# Inversion of Arctic dual-channel sound speed profile based on random airgun signal

Jinbao Weng[1,2,a], Yubo Qi[3], Yanming Yang[1,2], Hongtao Wen[1,2], Hongtao Zhou[1,2], Benqing Chen[1,2], Dewei Xu[1,2], Ruichao Xue[1,2], Caigao Zeng[1,2]

1. Laboratory of Ocean acoustics and Remote Sensing, Third Institute of Oceanography, Ministry of Natural Resources, Xiamen, Fujian 361005, China
2. Fujian Provincial Key Laboratory of Marine Physical and Geological Processes, Xiamen, Fujian 361005, China
3. State key laboratory of acoustics, Institute of Acoustics, Chinese Academy of Sciences, Beijing 100190, China

## ABSTRACT

For the unique dual-channel sound speed profiles of the Canadian Basin and the Chukchi Plateau in the Arctic, based on the propagation characteristics of refracted normal modes under dual-channel sound speed profiles, an inversion method using refracted normal modes for dual-channel sound speed profiles is proposed. This method proposes a dual-parameter representation method for dual-channel sound speed profiles, tailored to the characteristics of dual-channel sound speed profiles. A dispersion structure extraction method is proposed for the dispersion structure characteristics of refracted normal modes under dual-channel sound speed profiles. Combining the parameter representation method of sound speed profiles and the dispersion structure extraction method, an inversion method for dual-channel sound speed profiles is proposed. For the common horizontal variation of sound speed profiles in long-distance acoustic propagation, a method for inverting horizontally varying dual-channel sound speed profiles is proposed. Finally, this article verifies the effectiveness of the dual-channel sound speed profile inversion method using the Arctic low-frequency long-range acoustic propagation experiment. Compared with previous sound speed profile inversion methods, the method proposed in this article has the advantages of fewer inversion parameters and faster inversion speed. It can be implemented using only a single hydrophone passively receiving random air gun signals, and it also solves the inversion problem of horizontal variation of sound speed profiles. It has significant advantages such as low cost, easy deployment, and fast computation speed.

---

[a] Email: wengjinbao@tio.org.cn

## I. INTRODUCTION

In recent decades, the marine environment of the Arctic Ocean has undergone drastic changes. Notably, these changes include the warming of Arctic Ocean waters and the intrusion of warm Pacific water, resulting in a dual-channel sound speed profile in the Canadian Basin and the Chukchi Plateau. This dual-channel sound speed profile has garnered significant attention from marine acoustic scientists due to its remarkable enhancement effect on low-frequency, long-range sound propagation beneath Arctic ice. However, there is currently limited extensive and long-term observation of the dual-channel sound speed profile. From the limited measurement results at specific stations, it can be observed that the intensity of the dual-channel sound speed profile gradually decreases with increasing latitude. It essentially disappears at latitudes up to 82°N and longitudes up to 170°E. Additionally, the intensity of the dual-channel sound speed profile gradually increases from east to west. Nevertheless, there is currently no comprehensive observational data to support the distribution range and intensity variation patterns of the entire dual-channel sound speed profile, as well as its temporal variation patterns with seasons. Studying the spatiotemporal variation patterns of the dual-channel sound speed profile essentially involves investigating the spatiotemporal changes of the intrusion of warm Pacific water. Therefore, researching this issue holds significant oceanographic importance and is conducive to analyzing the overall changes in Arctic Ocean waters.

Over the past decade, the research on dual-channel sound speed profiles in the Arctic Ocean has garnered significant attention from global ocean acoustic experts. Multiple underwater acoustic propagation studies, as well as related underwater acoustic detection and positioning, and low-frequency long-range underwater acoustic communication research, have been conducted in areas with dual-channel sound speed profiles, yielding a series of theoretical and practical achievements. By confining the acoustic energy within the dual-channel water body, the dual-channel sound speed profile achieves no interaction with sea ice or the seabed, thus eliminating attenuation effects caused by sea ice and seabed. Changes in sea ice have no impact on this channel, theoretically resulting in only forward propagation loss of the acoustic field. This enables long-range propagation within the dual-channel depth, which is particularly beneficial for underwater acoustic detection and positioning, as well as underwater acoustic communication. Notably, a one-year acoustic experiment in the Canadian Basin observed that under the combined effects of the dual-channel sound speed profile and sea ice, the loss of fixed-point acoustic propagation fluctuated by more than 60dB. The biennial ICEX experiment in the Arctic achieved low-frequency long-range underwater acoustic detection and positioning, as well as underwater acoustic communication effects, utilizing dual channels to achieve low-frequency long-range underwater acoustic communication over hundreds of kilometers.

Given the crucial importance of the seawater sound speed profile (SSP) for underwater sound propagation, especially in the spatially and temporally complex Arctic Ocean, it is imperative to conduct extensive measurements to obtain a vast array of SSPs from multiple stations and across different seasons. However, due to the extensive ice coverage in the Arctic Ocean, it has been challenging to comprehensively measure the SSPs of most of the water bodies, particularly those in the central ice zone. Current methods for measuring SSPs under sea ice cover include ice-based buoy measurements (ITP), measurements during submarine navigation under ice, long-duration measurements at anchor-based stations, and multi-station field measurements using research vessels. Among these, ITP faces the challenge of difficult-to-fix measurement routes, submarine measurements incur high costs, anchor-based measurements can only obtain sound speeds at a single station, and extensive multi-station measurements using research vessels are costly. Therefore, there is an urgent need to develop new, low-cost methods for obtaining SSPs in the Arctic Ocean, and acoustic inversion methods merit further exploration as a significant approach for application in this region.

Currently, the inversion of seawater sound speed profiles based on acoustic methods has been carried out for decades, whether in shallow or deep sea environments. Especially in shallow water environments, due to the significant impact of shallow water sound speed profiles on sound field characteristics and long-distance propagation, many studies on the inversion of shallow water sound speed profiles have been conducted, including methods that utilize time-domain waveform of acoustic signals, sound field propagation loss, ambient noise field, ocean reverberation field, and other sound field characteristics to invert seawater sound speed profiles, achieving promising results. In deep sea environments, since the sound speed profile is mainly in the surface layer of seawater, it has a smaller impact on long-distance sound field propagation, mainly affecting the deep sea surface sound channel. When the surface sound speed is similar to that near the seabed, it has a greater impact on long distances. The main methods for seawater sound speed inversion in deep sea environments are based on sound field propagation loss and the multi-path arrival structure of sound rays. However, there is less research on the inversion of seawater sound speed profiles in the Arctic, mainly due to the difficulty of acoustic experiments in the Arctic and the scarcity of sound propagation data and seawater sound speed profile data. Therefore, it is urgent to develop seawater sound speed profiles in the Arctic, which is beneficial for the development of Arctic underwater acoustic research and related application research, especially the development of some low-cost acoustic inversion methods.

This article primarily focuses on the ubiquitous dual-channel sound speed profiles in the Canadian Basin of the Arctic Ocean and the Chukchi Plateau. Based on the theory of normal mode sound field, it analyzes the dispersion structure of acoustic signals under these profiles. Specifically, the dual-channel sound speed profiles lead to dispersion crossings, and the stronger the dual-channel effect, the more severe the dispersion crossing. Thus, the dual-channel sound speed profile can be estimated by matching the dispersion crossing structure. Given the numerous sound speed values that need to be inverted with depth for the dual-channel sound speed profiles, a method is proposed to characterize these profiles using two parameters based on their spatial distribution characteristics, enabling rapid inversion. Next, experimental data are utilized to analyze and verify the effectiveness of dual-channel sound speed profile inversion, considering both horizontally constant and horizontally varying scenarios. This article represents the first attempt to invert dual-channel sound speed profiles in the Arctic, successfully achieving the inversion of horizontally varying dual-channel sound speed profiles. Furthermore, it utilizes experimental data to validate the effectiveness of the rapid inversion algorithm for Arctic dual-channel sound speed profiles. This research has significant implications for studying dual-channel effects, underwater acoustic detection, underwater acoustic communication, and changes in Arctic water bodies.

## II. THEORYS

According to the normal mode theory, the underwater sound field can be represented as the superposition of a series of normal modes, with each normal mode possessing its own group velocity and eigenfunction vertical distribution. Consequently, a broadband pulse signal, after propagating over a long distance, will form a dispersion structure characterized by the arrival of normal modes at different times, particularly in shallow water waveguides. For the unique semi-waveguide sound speed profile beneath the Arctic ice, the sound speed profile is approximately composed of two linear approximations with different gradients. Within the surface layer of this sound speed profile, a refractive-like normal mode waveguide will form, exhibiting a dispersion structure that increases with time and frequency. For the distinct dual-channel sound speed profile in the Canadian Basin and the Chukchi Plateau, the nonlinearly increasing sound speed profile in the surface layer leads to the emergence of crossovers within the dispersion structure.

Based on the measured sound velocity profile of the Chukchi Plateau, acoustic signal simulation was conducted. The water depth was 3800m, and both the source depth and the receiving depth were 10m. The calculation frequency band ranged from 10Hz to 100Hz. By calculating the received complex sound pressure at different frequencies, the time-domain waveform of the acoustic signal was obtained through inverse

Fourier transform. Meanwhile, time-frequency analysis was performed on the simulated acoustic signal to analyze its normal mode dispersion structure. To be consistent with subsequent experiments, the propagation distances in this simulation included 200km, 300km, 400km, and 518km. The time-domain waveform of the simulated acoustic signal and the corresponding time-frequency analysis are shown in the figure below. From the simulation result diagram, it can be seen that the amplitude of the acoustic signal gradually increases with time. This is due to the small group velocity and small dispersion of normal modes in the high-frequency part, resulting in basically consistent arrival times and thus larger energy at the final arrival. The duration of the acoustic signal increases with the propagation distance, which is due to the difference in group velocity between the low-frequency and high-frequency parts of normal modes, causing the arrival time difference to increase with distance. From the time-frequency analysis of the acoustic signal, a clear dispersion crossing phenomenon can be observed, especially in the high-frequency part.

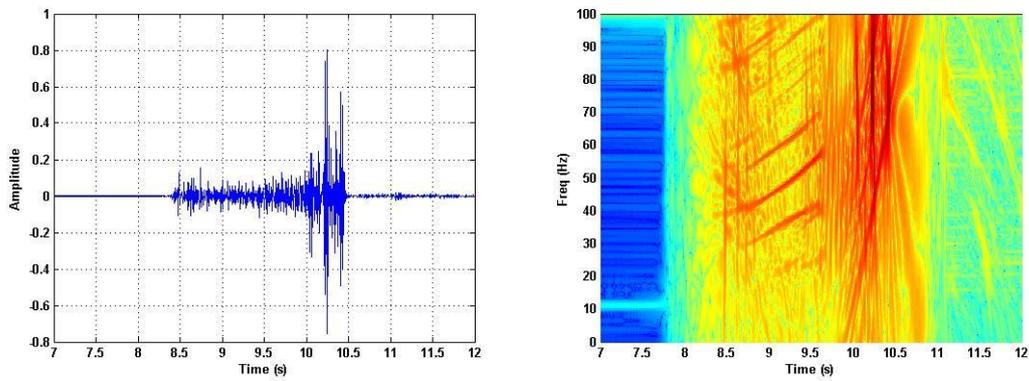

(a)

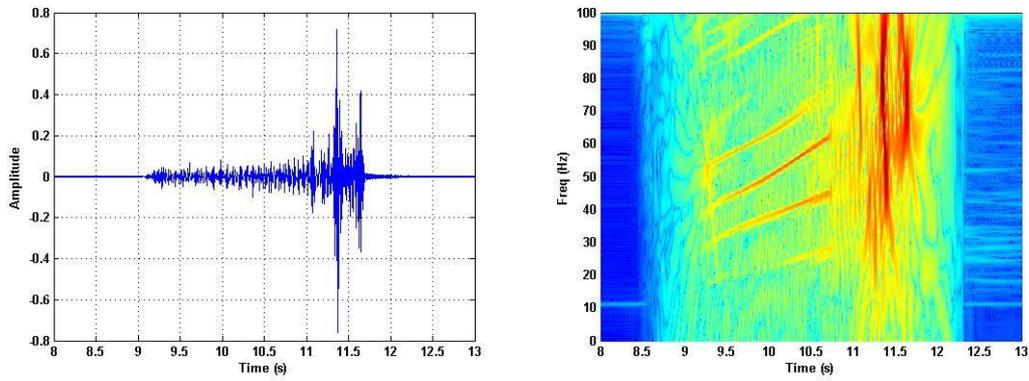

(b)

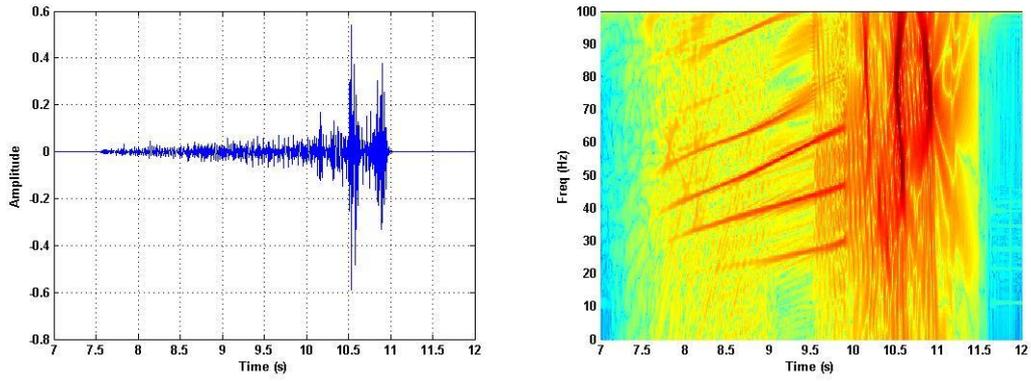

(c)

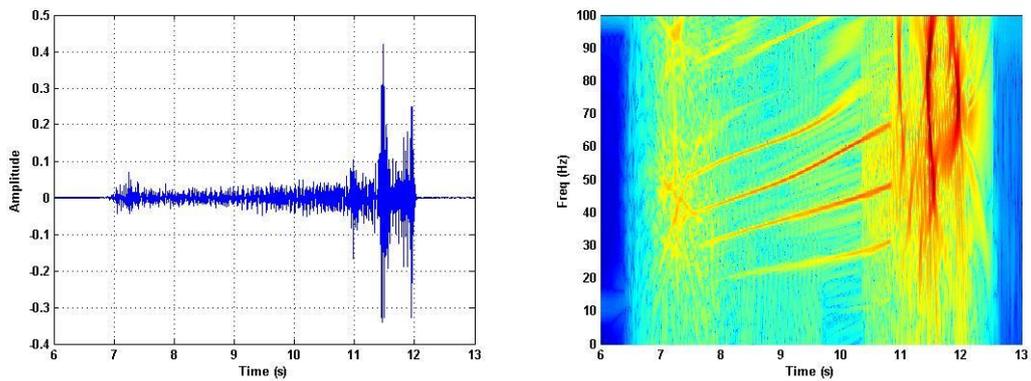

(d)

Figure 1 Simulated acoustic signal time-domain waveform and time-frequency analysis

Based on the measured two-channel sound speed profile, the group velocities of different normal modes at different frequencies were calculated using the normal mode acoustic propagation model. Subsequently, the dispersion curves at different propagation distances were calculated based on the propagation distance. The dispersion curves were compared with the simulated time-frequency analysis diagram of the acoustic signal, and the results are shown in the figure below. It can be seen from the figure that the dispersion curves calculated by the model are basically consistent with the dispersion structure of the simulated acoustic signal. In the high-frequency part of the normal modes, modes 1, 2, and 3 arrive in sequence, which is reversed from the order of modes 3, 2, and 1 in the low-frequency part of the normal modes, thus forming a dispersion crossover phenomenon. The dispersion crossover phenomenon is caused by the group velocity issues of different normal modes, and the more fundamental reason is the nonlinear increase in sound speed in the surface layer.

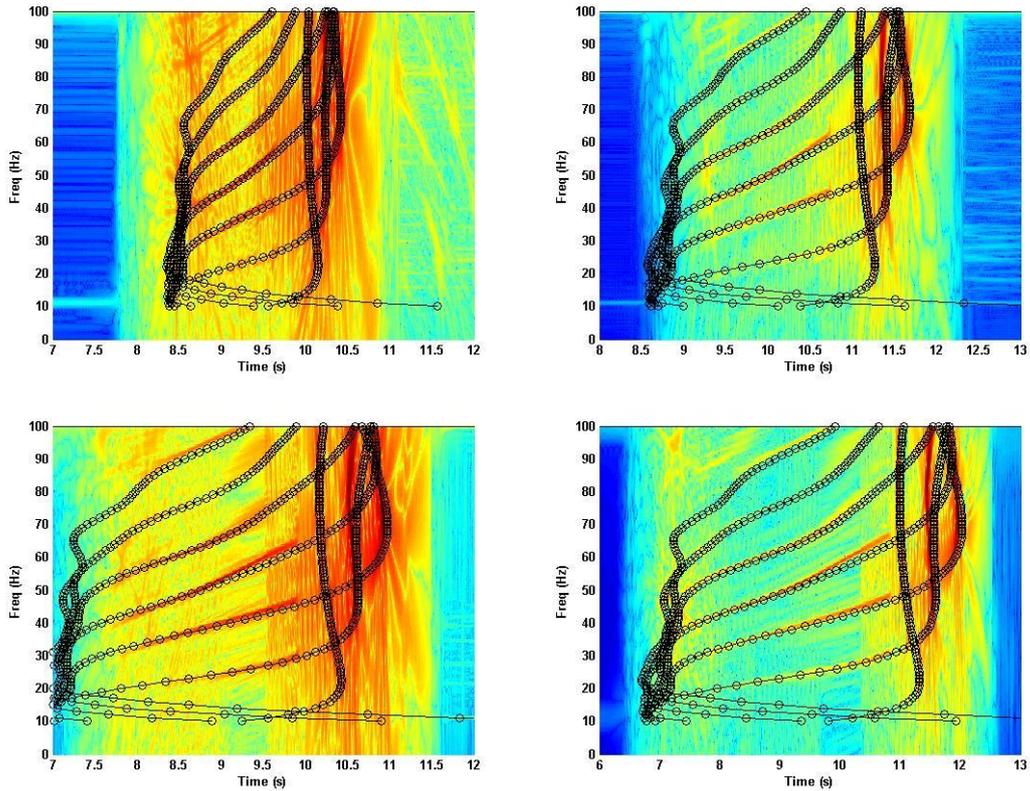

Figure 2 Time-frequency analysis of simulated acoustic signals and simulated dispersion curves

## III. A dual-parameter representation method for dual-channel SSP

For the dual-channel sound speed profiles in the Arctic waters, if the sound speed at each depth is directly inverted during the inversion process, the number of parameters to be inverted will reach hundreds. This will not only make the computation of the inversion work extremely large, but also require a huge amount of information for the inversion. Pure dispersion structures will not be able to meet the needs of this inversion method. Therefore, this article analyzes a large number of dual-channel sound speed profiles collected in the Chukchi Plateau and a large number of half-channel sound speed profiles collected in the central ice zone, conducts a comparative analysis of the two, and develops a dual-channel sound speed profile representation method that requires only two parameters. This method can greatly reduce the number of parameters to be inverted, reduce the computation and information required for the inversion process, and make dual-channel sound speed profile inversion a feasible solution.

## A. The difference between the dual-channel and single-channel SSP

This article compares the sound speed profiles collected at multiple stations in the Chukchi Plateau with the mono-channel sound speed profiles in the central ice zone, calculates the difference between the two, and plots the curves of the difference in sound speed profiles as a function of depth together. The results are shown in the figure. From the figure, it can be seen that the difference between the two is mainly between 0m and 400m, divided into positive and negative difference segments. The dividing point between the positive and negative difference segments is concentrated between 100m and 150m, and the maximum values of positive and negative differences are basically consistent, both around 10m/s. The maximum positive and negative differences for each sound speed profile are calculated, and the results are shown in the figure. From the figure, it can be seen that the maximum positive and negative differences for each sound speed profile are basically consistent.

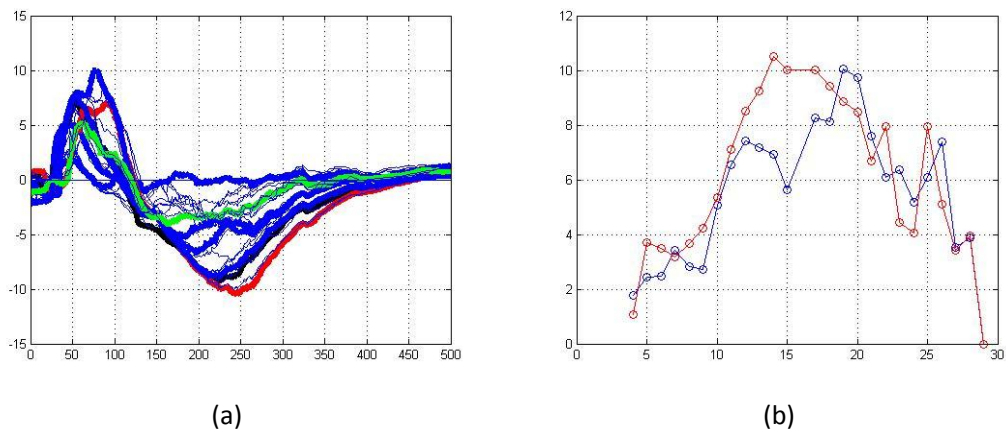

(a)                                                      (b)

Figure 3 The differences between the dual-channel sound speed profiles of the Chukchi Plateau in the Arctic and the sound speed profiles of the central ice zone, (a) the curve showing the variation of sound speed difference between 26 measured dual-channel sound speed profiles of the Chukchi Plateau and the sound speed profiles of the central ice zone with depth, (b) the comparison curve of the maximum and minimum values (absolute values) of the 26 sound speed difference curves

## B. A dual-parameter representation method for dual-channel SSP

Based on the characteristics of the differences between the Arctic dual-channel sound speed profile and the half-channel sound speed profile, this article proposes an approximate formula expression for the dual-channel sound speed profile. The Arctic dual-channel sound speed profile can be represented as the sound speed profile of the central ice zone in the Arctic plus four linear functions. The linear functions can be represented by two parameters: one is the length, and the other is the amplitude. The peak values of the quartic functions are consistent, but the lengths are different. Therefore, these two parameters can represent this difference: one is the channel intensity, and the other is the channel width. These two parameters are the dual-channel intensity, which is the upper and lower peak values, and the dual-channel width, which is the length of the first linear function, representing the curve of sound speed difference with depth. The half-period of the first linear function ranges from 30m to the channel width. The period of the second linear function ranges from the remaining depth to 400m. Therefore, the sound speed width and intensity are set as unknown variables. The piecewise linear function could be expressed as

$$c(z,I,W) = c_0(z) + \begin{cases} -1, z \leq 30 \\ (z-31) \bullet I/W, 31 \leq z \leq 31+W \\ I-(z-(31+W)) \bullet I/2W, 31+W+1 \leq z \leq 31+3W \\ -(z-(31+3W)) \bullet I/((400-(31+3W))/3), 31+3W+1 \leq z \leq 31+3W+(400-(31+3W))/3 \\ -(400-z) \bullet I/(2 \bullet (400-(31+3W))/3), 31+3W+(400-(31+3W))/3+1 \leq z \leq 400 \\ 0, 401 \leq z \leq 501 \end{cases} \quad (1)$$

Of the two parameters, one is labeled I and the other W, representing 1/3 of the dual-channel intensity and dual-channel width, respectively.

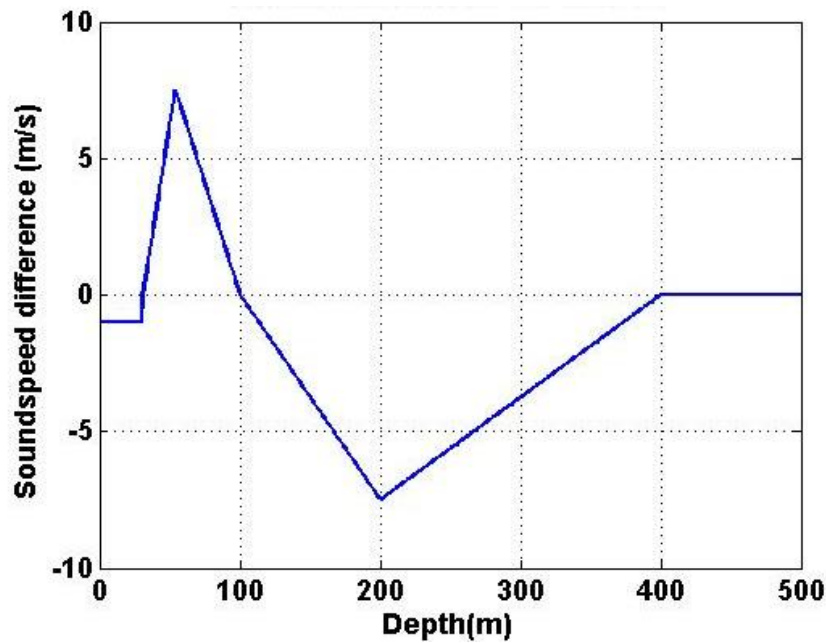

Figure 4 Approximate representation curve of the sound speed difference between the two-channel sound speed profile and the sound speed profile in the central ice zone

In this paper, we established a cost function to find the optimal dual parameters for representing the two-channel sound velocity profile

$$\text{Cos}tFunction(I,W) = \sum_z \left| c(z,I,W) - c_m(z) \right|^2 \qquad (2)$$

For each specific two-channel sound speed profile, the corresponding two parameter values can be found through a cost function. As shown in the figure below, when the sound speed profile is as depicted on the right, the dual parameters can be determined using the cost function. The ambiguity of the cost function is illustrated in the figure below, where the optimal value is achieved with a channel intensity of 7.5 m/s and a channel width of 69 m. By substituting the found optimal values into the expression function, the corresponding two-channel sound speed profile is obtained. When compared with the original two-channel sound speed profile, the results are essentially identical, as shown in the figure below. Therefore, this example demonstrates the effectiveness of representing two-channel sound speed profiles using dual parameters.

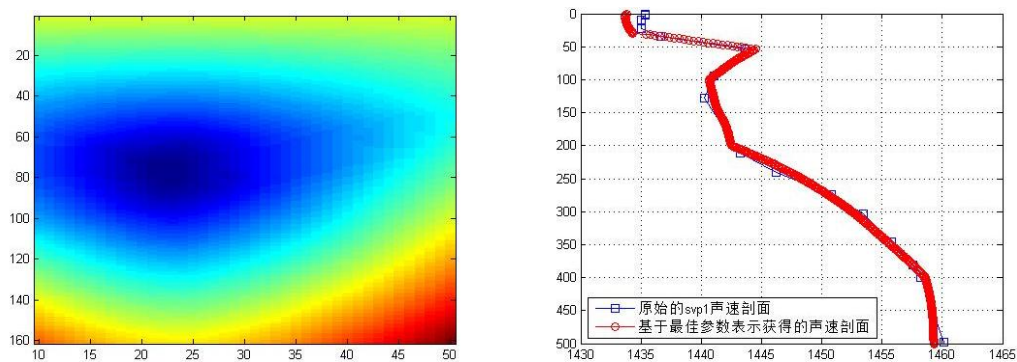

Figure 5 The optimal dual-parameter found under the dual-parameter representation method, and the comparison between the sound speed profile under the optimal dual-parameter representation and the original sound speed profile

## IV. A fast inversion method for dual-channel SSP

Due to the phenomenon of broadband acoustic signal dispersion and cross-over caused by the Arctic dual-channel sound speed profile, and the fact that the dual-channel sound speed profile can be represented by two parameters, the dual-channel sound speed profile can be quickly inverted by matching the cross-dispersion structure of broadband acoustic signals obtained by a single hydrophone. For these reasons, this article proposes a single-hydrophone inversion method for the Arctic dual-channel sound speed profile.

Since the inversion method proposed in this article is achieved by matching dispersion structures, which are essentially caused by different group velocities of different frequencies and different normal modes, it is necessary to calculate the group velocities of different normal modes at different frequencies under all sound speed profiles in advance before inverting the two-channel sound speed profile. Specifically, common normal mode sound field calculation models, such as Kraken and Orca, can be utilized to calculate the group velocities of different normal modes at different frequencies under different two-parameter representations for different two-channel sound speed profiles within the working frequency band (the working frequency band in this article is 10Hz to 100Hz). Then, based on the propagation distance of acoustic signals, the dispersion curves under different sound speed profiles can be calculated, which can be used as the comparison between the copied field and the measured cross-dispersion curve.

For low-frequency broadband pulse acoustic signals that have propagated over long distances, the acoustic signals will form a cross-type dispersion structure in time-frequency analysis under the influence of a two-channel sound speed profile. Due to the presence of crossings in the dispersion curve, it is impossible to use warping transformation to transform each normal mode into a single-frequency signal for modal separation in one go. To address this issue, this paper proposes a segmented processing method to extract the cross-type dispersion curve. The broadband acoustic signals received by a single hydrophone are divided into non-crossing segments and crossing segments in the time domain. For the non-crossing segments, warping transformation is used for modal separation and dispersion curve extraction. For the crossing segments, energy extrema points at different frequencies are used to extract the dispersion curve. Finally, the two are combined to obtain a complete cross-type dispersion curve. Based on this method, experimental data will be used to verify its effectiveness in the future.

By calculating the group velocity of normal modes under different sound speed profiles with varying dual-channel intensities and widths, and combining this with the propagation distance, we obtain dispersion curves for different dual-channel intensities and widths. By comparing the simulated dispersion curves with the experimental ones, when the two achieve the best match, the sound speed profile under the current dual-channel intensity and width is the optimal inversion result.

$$\mathrm{Cos}tFunction(I,W) = \sum |Disp(I,W) - Disp_m|^2 \tag{3}$$

Theoretically, the dispersion structure undergoes significant changes with propagation distance, making it possible to invert the source distance based on the dispersion structure. Therefore, this article simultaneously inverts the source distance and two parameters of the dual-channel, by matching the dispersion curve relative structure, with the reference point as the zero point, and comparing the arrival times under different modes and frequencies. When the two match best, the current source distance, dual-channel width, and intensity are selected as the inversion results.

For the two-channel sound speed profile inversion method in horizontally varying marine environments, the inversion is performed segment by segment through the piecewise horizontal invariance approximation to obtain the horizontally varying results. This segmented inversion method requires a premise to be established, which is that the distance of the sound source is fixed, and the distances of different segments are also fixed. Simultaneous inversion iterations are carried out, starting with the first segment, then the second segment, and finally obtaining the inversion results for each segment.

## V. Introduction to an acoustic propagation experiment

To verify the effectiveness of the two-channel sound speed profile inversion method, this article utilizes random air gun sound signals obtained from an Arctic ambient noise measurement experiment to conduct the verification of the inversion method. We deployed long-term marine ambient noise observation equipment at the Chukchi Plateau. After the equipment was recovered, the received noise signals were analyzed, and clear continuous broadband pulse sound signals were observed from the signals. At the same time, we searched for information on ships working in the experimental sea area based on the time of the received sound signals. Finally, through multiple angles of verification, it was confirmed that the received sound signals were from the seismic exploration experiment conducted by the experimental vessel Sikuliaq at the Chukchi Plateau, including sound signal simulation verification at multiple different distance stations. The specific sound signal verification work can be referred to in the sound signal processing and analysis section. Next, this article will detail the working route of the experimental vessel where the sound source is located and the environmental information of the working sea area, followed by the processing and analysis of the received sound signals.

## A. Sound source and receiving device

Based on the plan book published on the Internet, the proposed seismic survey would be expected to last for ~45 days, including ~30 days of seismic operations, ~8 days of transit to and from the survey area, and 7 days for equipment deployment/recovery. R/V Sikuliaq would likely leave out of and return to port in Nome, AK, during summer (August/September) 2021. During the surveys, R/V Sikuliaq would tow up to 6 G-airguns. The array would be towed at a depth of 9 m, and the shot interval would be 35 m (~15 s) during MCS surveys, 139 m (60 s) during refraction surveys.

The MMSI of the SIKULIAQ vessel is 338417000. Based on the satellite-derived information regarding the vessel's time and heading trajectory, its actual navigation trajectory does not align with the planned one, and the specific reason remains unclear. However, this does not affect the specific research content of this article. Through online search, we obtained the environmental impact assessment report for the seismic exploration experiment conducted by the test vessel Sikuliaq in the Chukchi Plateau, providing partial information about the seismic experiment.

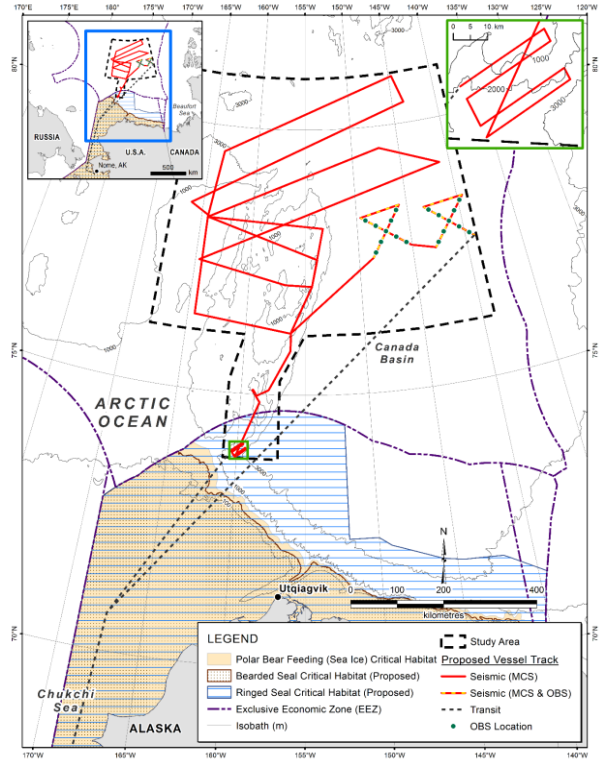

Figure 6 The planned navigation route for the seismic exploration experiment in the Chukchi Plateau of the Arctic based on the published environmental assessment report of Surveys

We obtained the actual navigation route of the test vessel using the AIS information from the purchased Sikuliaq vessel, as shown in the figure below. It can be seen from the figure that the actual navigation route does not align with the planned route, but this discrepancy will not affect the dual-channel sound speed profile inversion work.

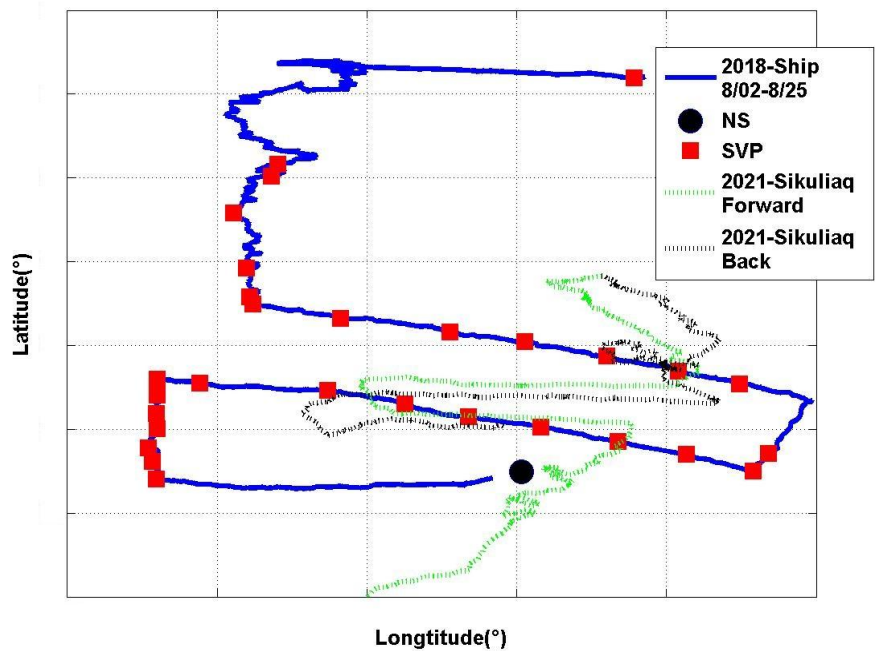

Figure 7 Actual navigation route and sound velocity profile measurement stations for seismic exploration experiments in the Chukchi Plateau of the Arctic

Below is an overview of the receiving equipment, specifically the moored buoys on the Chukchi Plateau. The acoustic signal receiving equipment was deployed in the waters of the Chukchi Plateau in the summer of 2020. The equipment began operating at 2020/08/04/010000 and ended at 2021/09/08/080000. The sampling rate was 16k, with data being collected for 2 minutes every hour, starting at the beginning of each hour. A total of six vertical receiving elements were deployed this time, numbered 01, 03, 04, 06, 07, and 08. No synchronization measures were taken between the six elements, with two elements forming a group. Among them, three TDs were deployed, numbered 082822, 202971, and 202982, with depths of 52m, 882m, and 1662m, respectively. Currently, the data from element 01 is primarily used for analysis because it is located at the surface depth, and the data from element 03 is basically consistent with that of element 01.

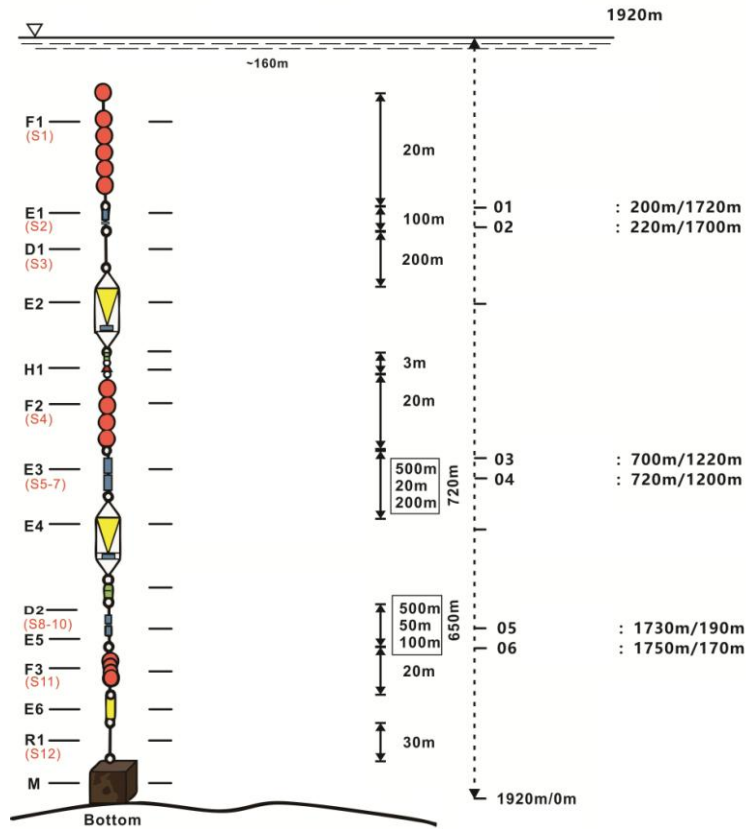

Figure 8 Acoustic buoys deployed in the Chukchi Plateau during China's 11th Arctic scientific expedition in 2020

## B. Ocean environment

Next, we will introduce marine environmental information. Firstly, we will discuss the seabed topography (based on the seabed topography of the experimental sea area obtained from ETOP), as shown in the figure below. The seabed topography of the experimental sea area is complex, varying from several hundred meters to several kilometers. Therefore, how to avoid the impact of seabed topography changes on sound speed profile inversion is a problem that needs to be considered. This article mainly uses acoustic signals within the deep-sea surface channel to invert the sound speed profile, so the impact of seabed topography changes is minimal.

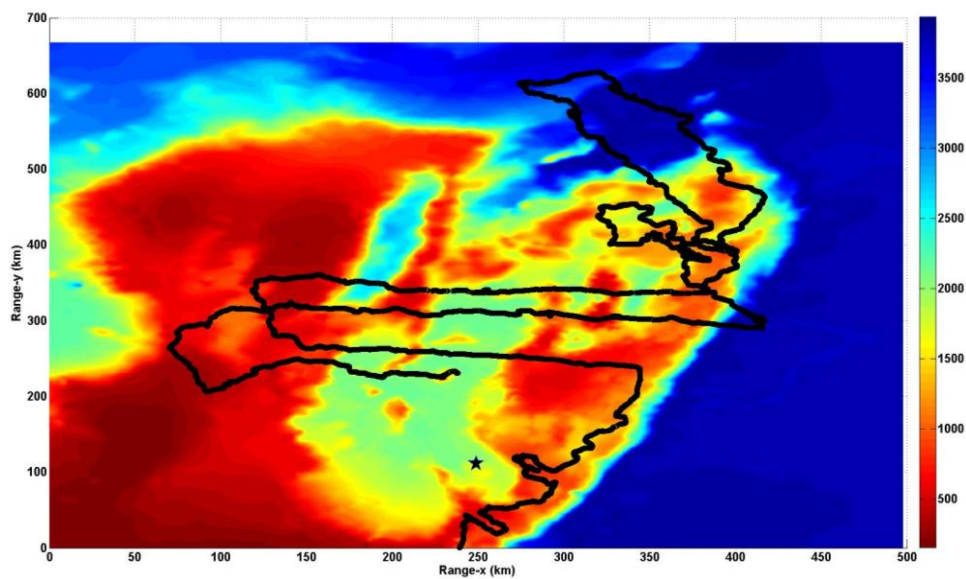

Figure 9 The seabed topography of the experimental sea area obtained based on the ETOP database

Below is an introduction to the sound speed profile (SSP), which describes the spatial distribution of the SSP based on experimental data. According to the actual sound propagation measurement lines and station information, it can be seen that the sound speed profile information of the air gun test area was collected at three SSP measurement stations on the Chukchi Plateau, spanning from east to west, from west to east, and from south to north. The following figure shows the variation process of the sound speed profile along these three measurement lines. It can be seen from the figure that the intensity of the dual-channel sound speed profile gradually increases from east to west and gradually decreases from south to north, and the dual-channel sound speed profile basically disappears at the station located at 82 °N and 168 °W.

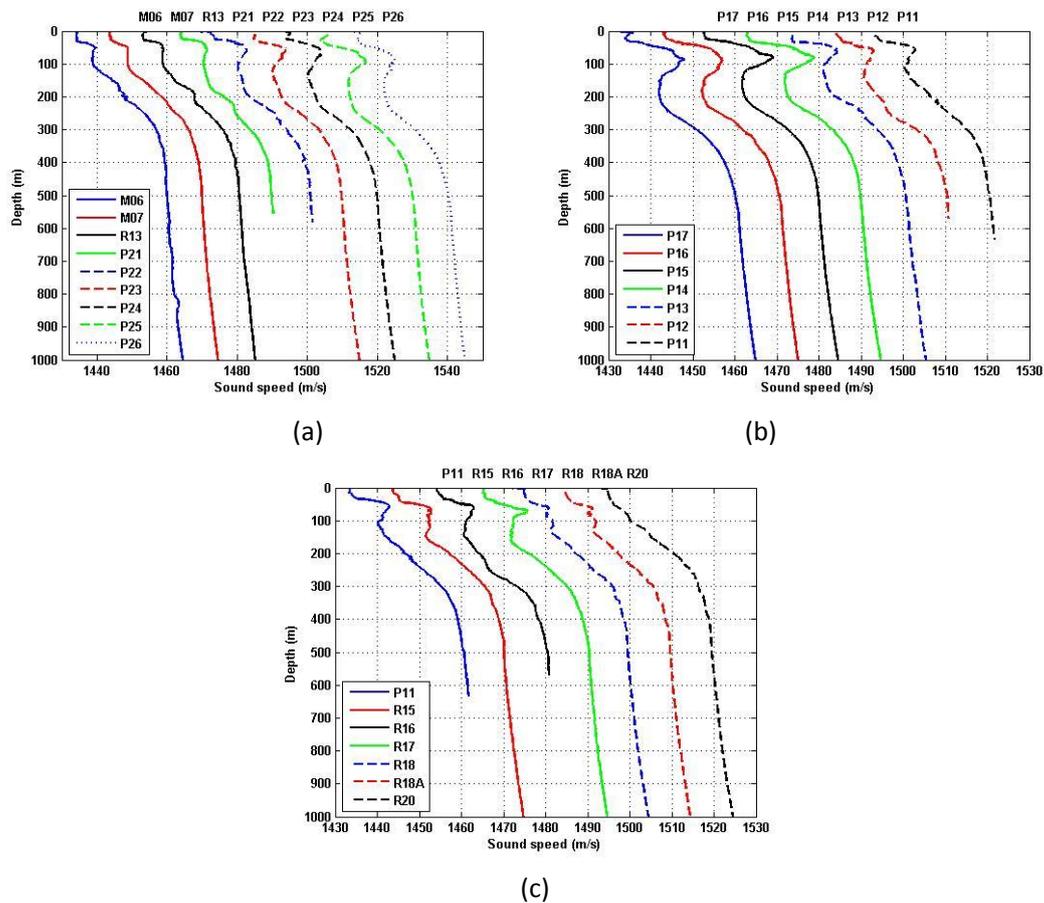

(a)

(b)

(c)

Figure 10 Spatial variation of the sound velocity profile measured in the experimental sea area

## C. Receiving sound signals

Next, we will introduce the received acoustic signals. Based on the AIS information of the ship, we selected four typical acoustic propagation measurement lines for analysis of the received acoustic signals and two-dimensional sound field. The figure below shows the four selected stations and their corresponding propagation distance information.

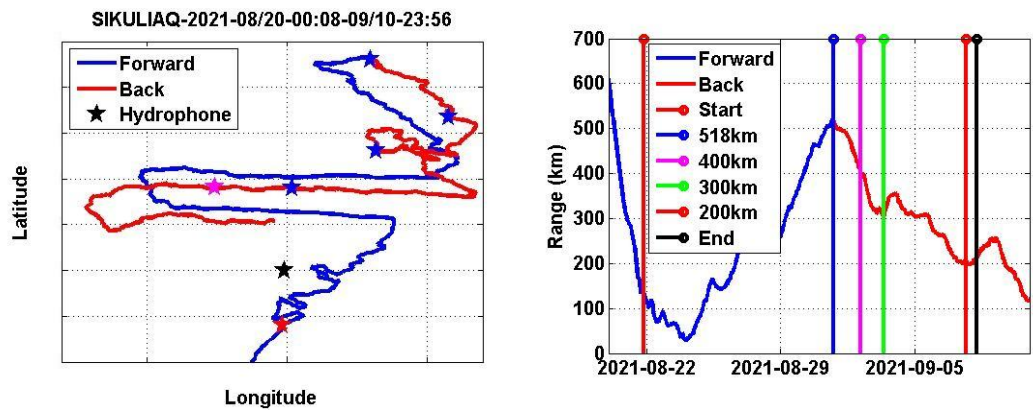

(a) (b)

Figure 11 The sound source station locations for four typical sound propagation measurement lines, (a) information on the four station locations, (b) information on the distances between the four stations

Based on the seabed topography along the sound propagation paths of the four stations obtained from ETOP, a two-dimensional sound field simulation was conducted to further confirm the distribution of the sound field. It can be observed that although the seabed topography exhibits significant horizontal variations, it does not affect the long-distance propagation of acoustic signals in the surface layer. In other words, it does not affect the sampling of surface water for the acoustic signals used to invert the sound speed profile. Therefore, for the method of inverting the two-channel sound speed profile proposed in this article, the horizontal variations in seabed topography do not affect the effectiveness of the inversion method.

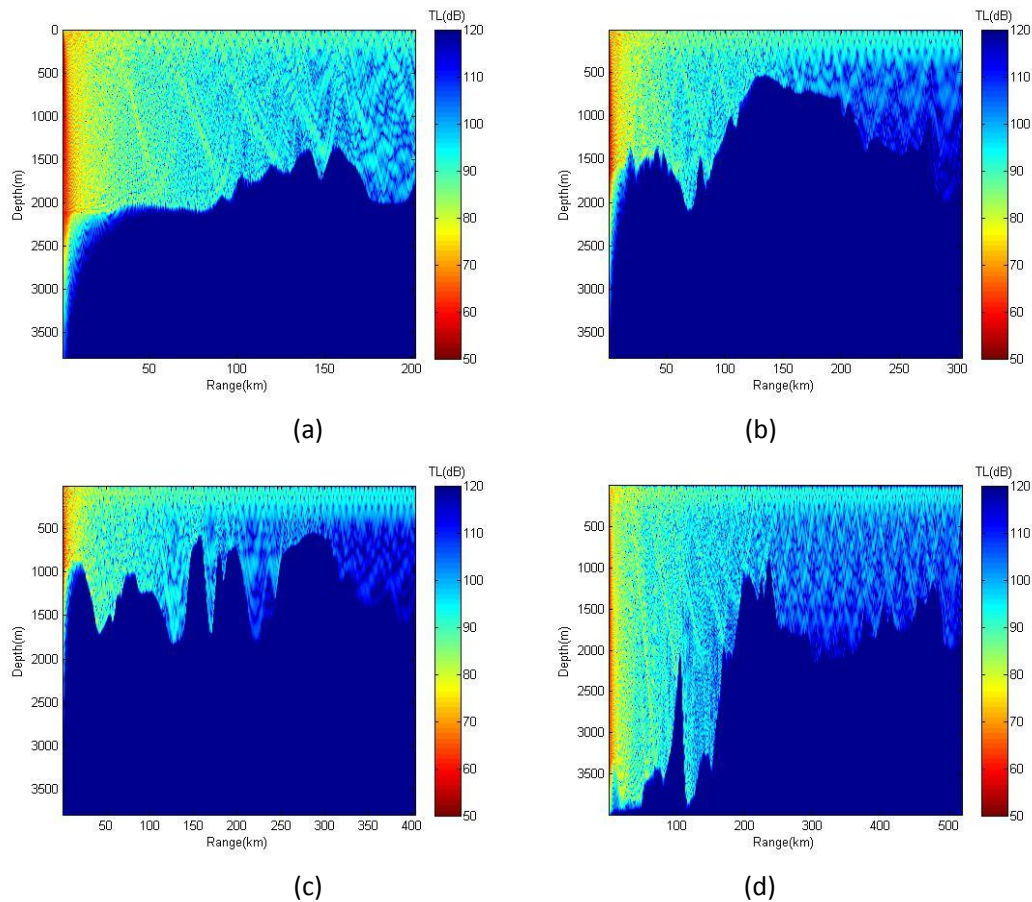

(a) (b)

(c) (d)

Figure 12 Simulation results of two-dimensional sound field on four sound propagation paths, (a) 200km, (b) 300km, (c) 400km, (d) 518km

Next, time-domain waveform analysis and time-frequency analysis were conducted on received acoustic signals from four different distances. Combining time-domain waveform and time-frequency analysis, it can be observed that the seabed topography can block the arrival of multi-path acoustic rays that have undergone deep inversion, as well as the low-frequency part of normal modes, but not the high-frequency part of normal modes. Therefore, the seabed topography has no impact on the inversion of the two-channel sound speed profile in the surface water. At the same time, from the time-frequency analysis of acoustic signals from four different distances, it can be seen that the long-distance broadband acoustic signals in this sea area generally exhibit frequency dispersion crossover, which is closely related to the fact that this sea area is a two-channel sea area. Additionally, since the acoustic source is an air gun with a very shallow depth of only about ten meters, according to the characteristics of the eigenfunctions of the Arctic surface refracted normal modes, a very shallow acoustic source depth can excite a relatively complete frequency dispersion structure, especially the high-frequency part of normal modes, which is the frequency dispersion crossover part. Therefore, using a low-frequency broadband shallow-depth air gun acoustic source has significant advantages in the inversion of the sound speed profile in the Arctic surface water.

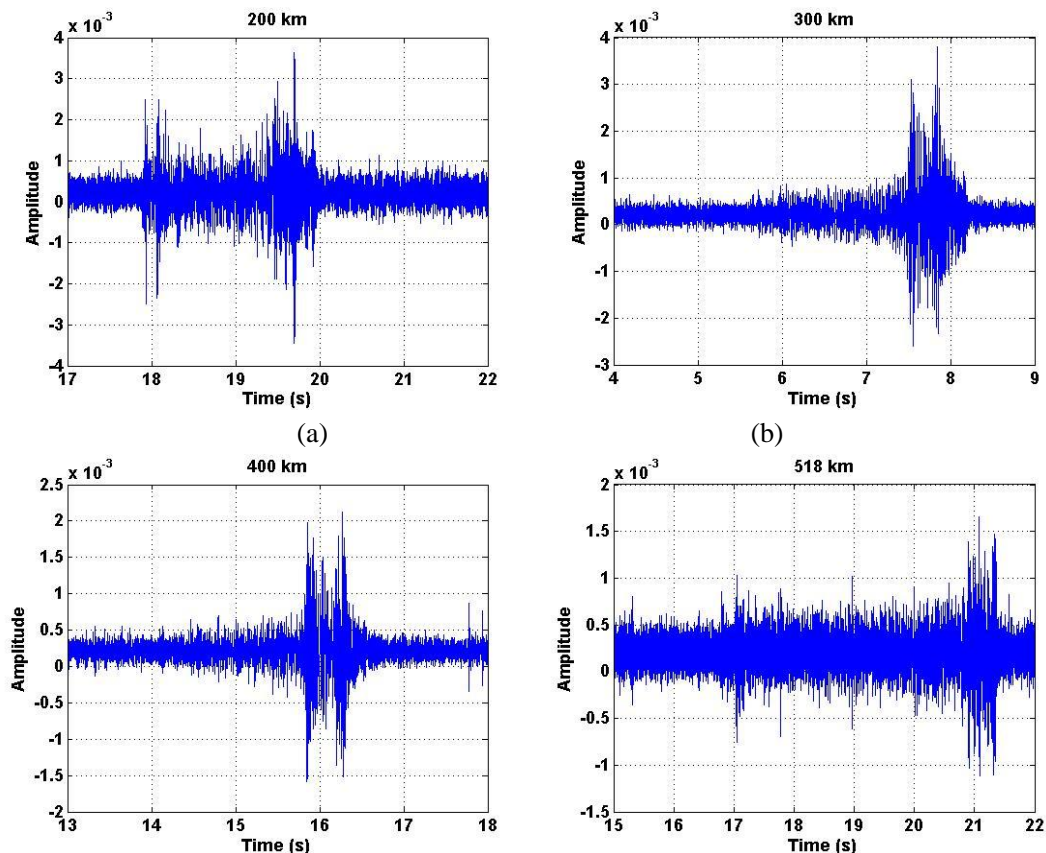

(a) (b)

(c) (d)

Figure 13 Time-domain waveform of acoustical signals in four typical propagation examples received during the experiment

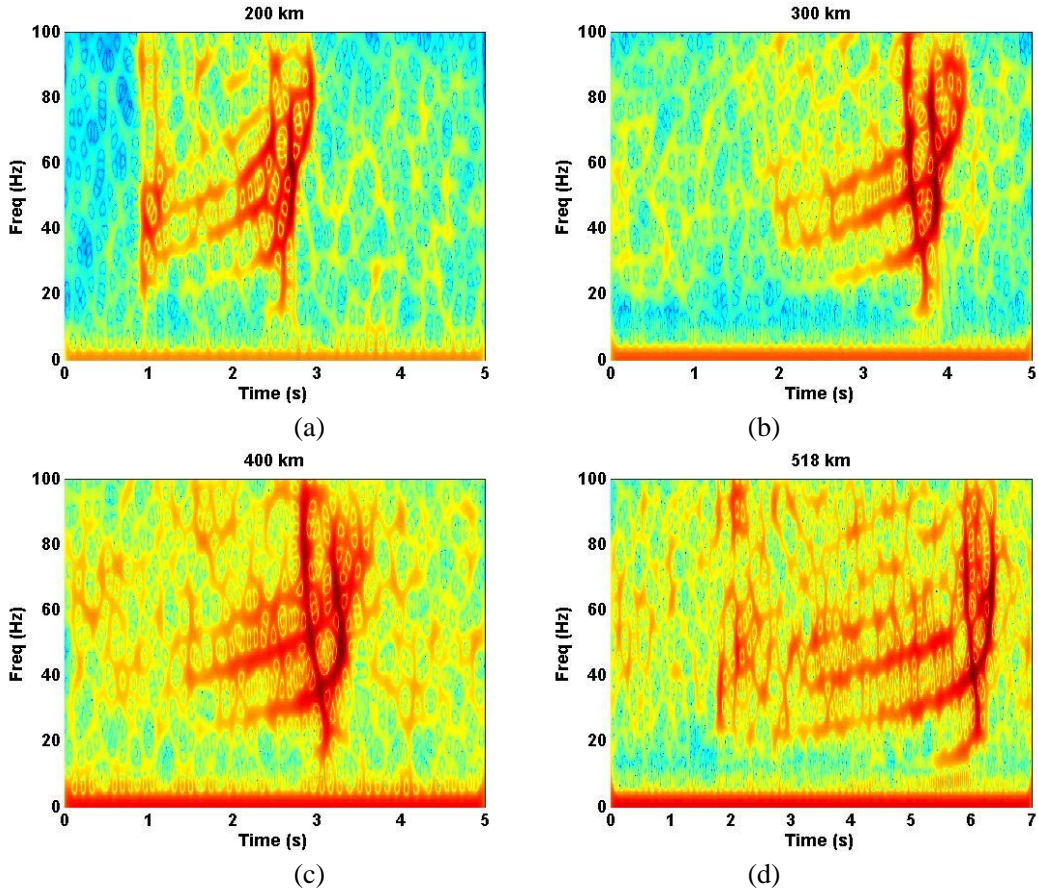

(c) (d)

Figure 14 Time-frequency analysis of acoustic signals in four typical propagation examples received during the experiment

## D. Extracted normal mode dispersion curve

Based on the broadband air gun acoustic signals collected from experiments at long distances, we processed the received acoustic signals and extracted the dispersion curves of the broadband acoustic signals for further use in matching the dispersion curves to invert the two-channel sound speed profile. According to the dispersion structure characteristics of refracted-like normal modes in the surface layer of the Arctic deep sea, the broadband acoustic signals at long distances were divided into two parts in the time domain for dispersion curve extraction. As shown in the figure below, the low-frequency part of the dispersion structure was separated into modes and the dispersion curve was extracted using the warping transformation of refractive index normal modes. For the high-frequency part of the dispersion structure, dispersion curves were extracted using energy extrema points at different frequencies. The combination of the two methods yielded a complete dispersion curve. It can be seen from the figure that the extracted dispersion curve is consistent with the dispersion structure in the time-frequency analysis of the background. From the extracted dispersion curve, it is also clear that in the dispersion crossing part, the first mode arrives first, followed by the second mode, and finally the third mode, which is opposite to the order in the low-frequency part. In the low-frequency part, there is no crossing phenomenon in the dispersion curve, and the arrival structure of each normal mode is consistent with the dispersion structure of the half-waveguide sound speed profile. Therefore, using only the dispersion curve in the low-frequency part cannot invert the two-channel sound speed profile. In this paper, we selected the strongest frequency point in each received acoustic signal as a reference point, and then used this point as a reference for the rest. The two selected reference points are 20Hz for the first mode and 28Hz for the second mode, both of which have strong modal energy. Since the absolute propagation time is unknown, inversion needs to be performed by matching relative dispersion curves. At this time, a reference point needs to be selected to simultaneously invert the distance and sound speed profile based on the experimental dispersion structure.

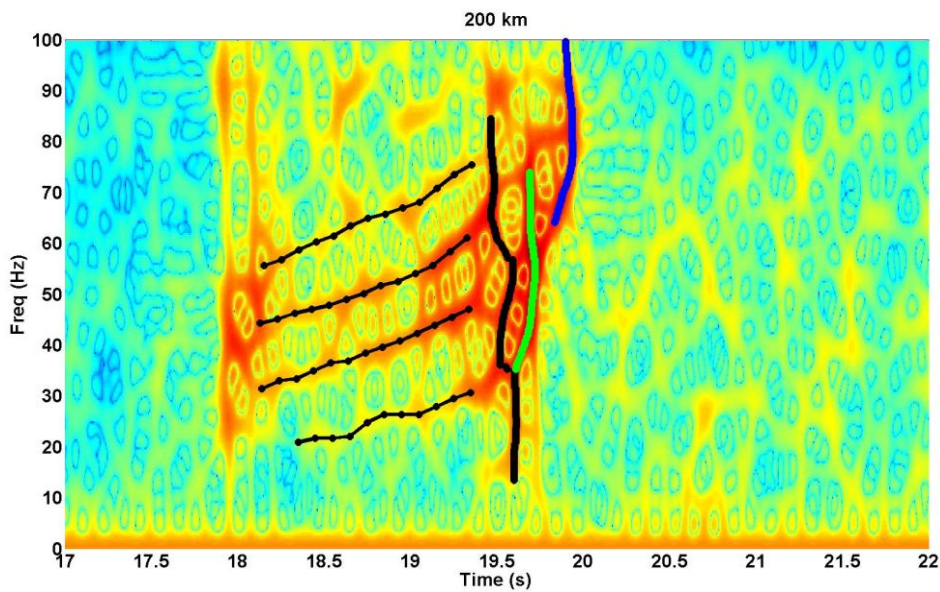

(a)

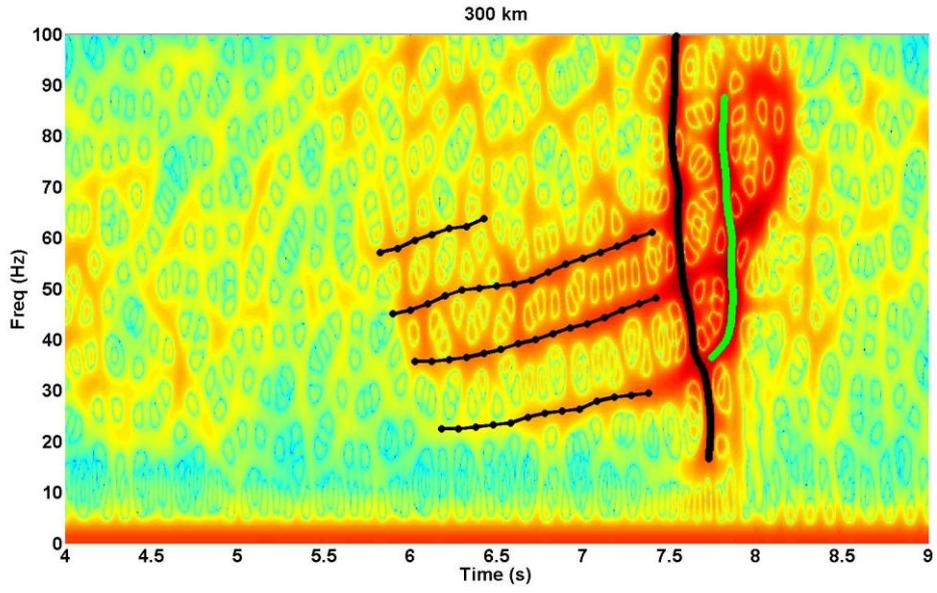

(b)

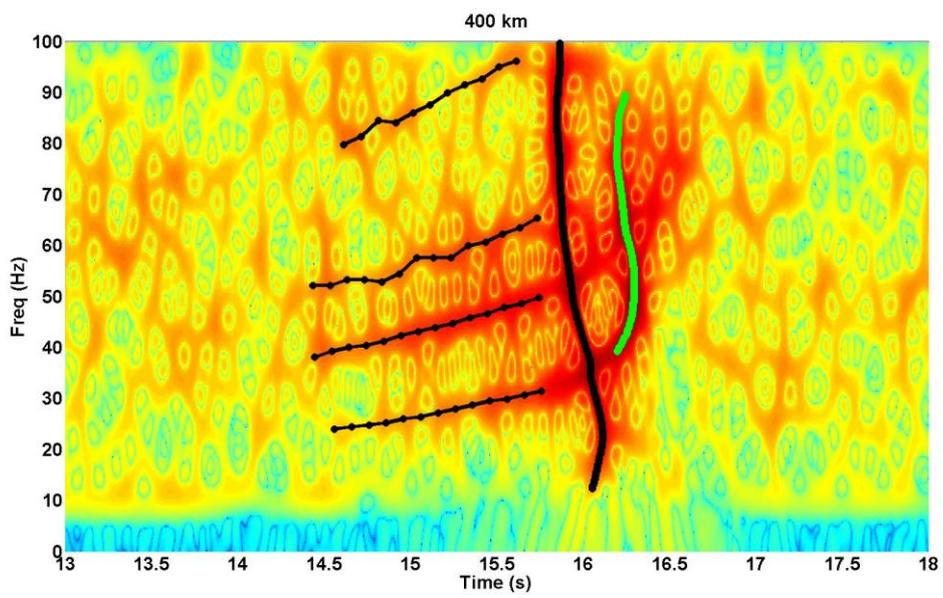

(c)

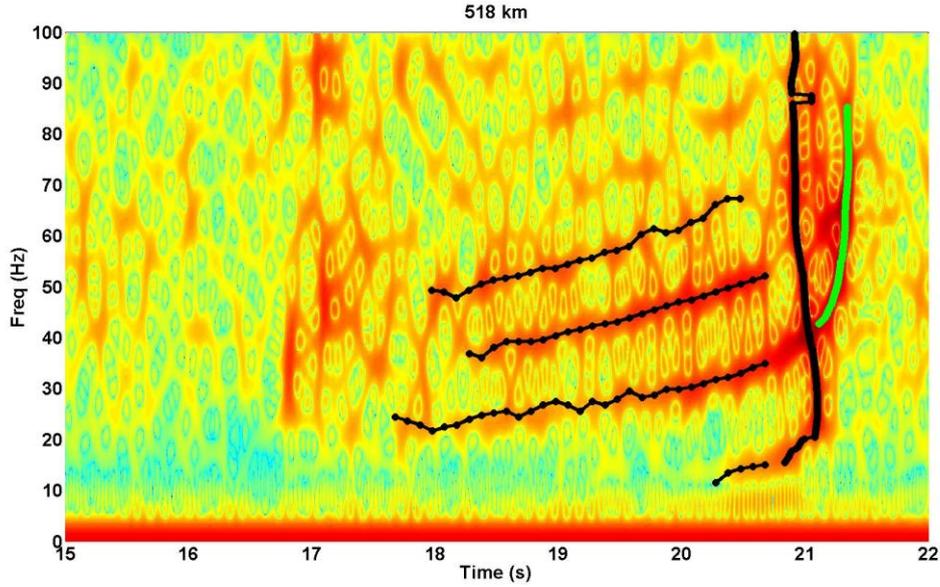

(d)

Figure 15 The dispersion curves of normal modes extracted from received acoustic signals, (a) 200km, (b) 300km, (c) 400km, (d) 518km

## VI. Experimental inversion results of dual-channel SSP

After extracting the measured dispersion curves from experimental data, this article conducted an experimental inversion of the sound speed profile for the Arctic dual-channel experiment. First, we introduce the dual-channel sound speed profile inversion result in a horizontally invariant environment based on dispersion curve inversion. Then, we present the dual-channel sound speed profile inversion result in a horizontally varying environment based on dispersion curve inversion. To verify the effectiveness of the Arctic dual-channel sound speed profile inversion method proposed in this article, we conduct dual-channel sound speed profile inversion using the dispersion curves of acoustic signals extracted at four different distances. To consider the application prospects of the method, this article first assumes that inversion is conducted without knowing the sound source distance, simultaneously inverting both the sound source distance and the sound speed profile. Subsequently, we conduct sound speed profile inversion using known sound source distances, and compare it with the former to analyze the influence of sound source distance information. Afterwards, we utilize dispersion curves of acoustic signals at different distances to demonstrate the effectiveness of the method for inverting the horizontally varying dual-channel sound speed profile under horizontally varying conditions.

## A. Inversion result in range-independent environment

Below are the inversion results of the two-channel sound speed profile in a horizontally uniform ocean environment. Assuming the horizontal distance between the sound source and the receiver is unknown, the distance is treated as an unknown parameter for joint inversion. First, we present the inversion results of the sound source distance and the two-channel sound speed profile under the condition of unknown sound source distance. Inversions are conducted for acoustic signals with four different propagation distances. Of course, the inversion results for these four different distance signals are all average sound speed profiles over the propagation distance. The following four figures show the inversion results for the four different distances. From the inversion results of the 200km acoustic signal, it can be seen that the inversion distance of 215km is close to the actual distance of 200km. The ambiguity diagram of the inverted sound speed profile dual parameters has a very obvious local optimum. The dispersion structure of the acoustic signal calculated based on the inverted sound speed profile is basically consistent with the measured dispersion curve, especially for the intersecting dispersion curves No.1, No.2, and No.3. Finally, a comparison between the inverted sound speed profile and the measured sound speed profile shows that they are relatively consistent. The 200km acoustic propagation sea area has a strong two-channel sound speed profile, and the inversion results are in good agreement with the measured two-channel sound speed profile. Based on the four results in the figure, including the inversion distance, dual parameter ambiguity diagram, dispersion curve comparison, and sound speed profile comparison, it can be seen that the two-channel sound speed profile inversion method matching the dispersion curve proposed in this article is relatively effective.

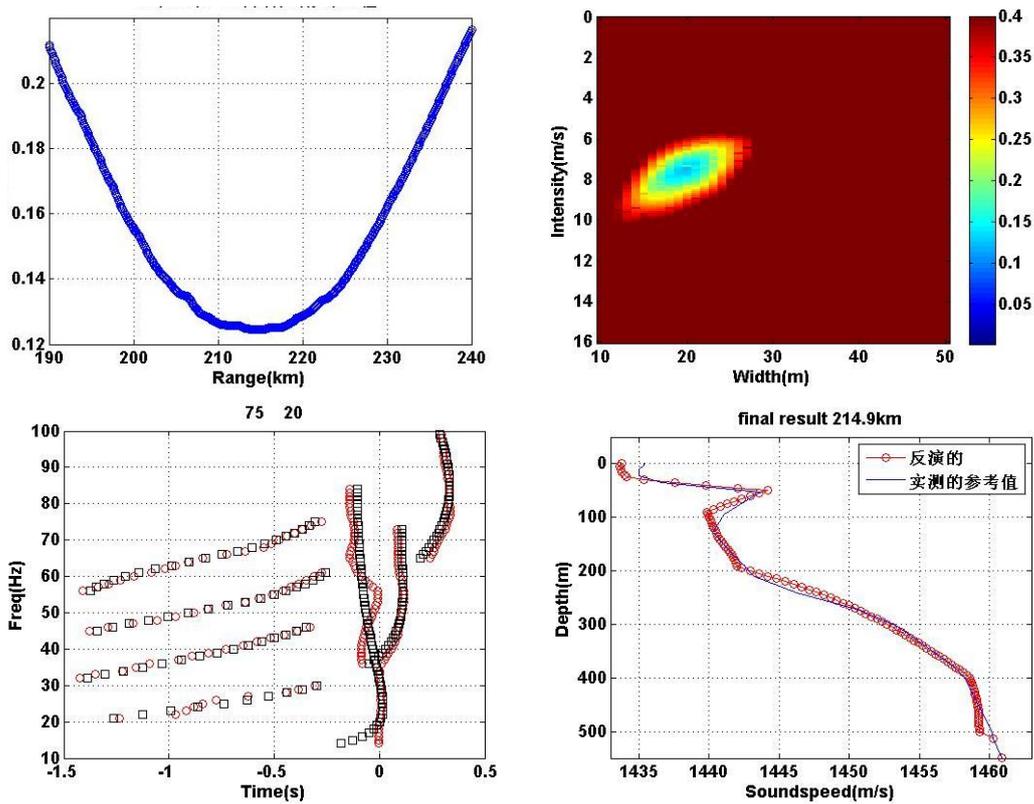

Figure 16 Cost function ambiguity map and sound velocity profile for 200km signal propagation inversion

Next, we performed sound speed profile inversion on acoustic signals from three other distances. The following three figures present the inversion results for these three distances. From the figures, it can be observed that the inversion results are similar to those of the first distance, indicating errors in the sound source distance inversion results. The sound speed profile has a very obvious local optimal solution for the dual-parameter ambiguity. The dispersion curve obtained from the inverted sound speed profile is relatively consistent with the measured dispersion curve, and the inverted sound speed profile is also consistent with the measured sound speed profile. Therefore, the method of using random air gun signals to invert the dual-channel sound speed profile in this article is effective.

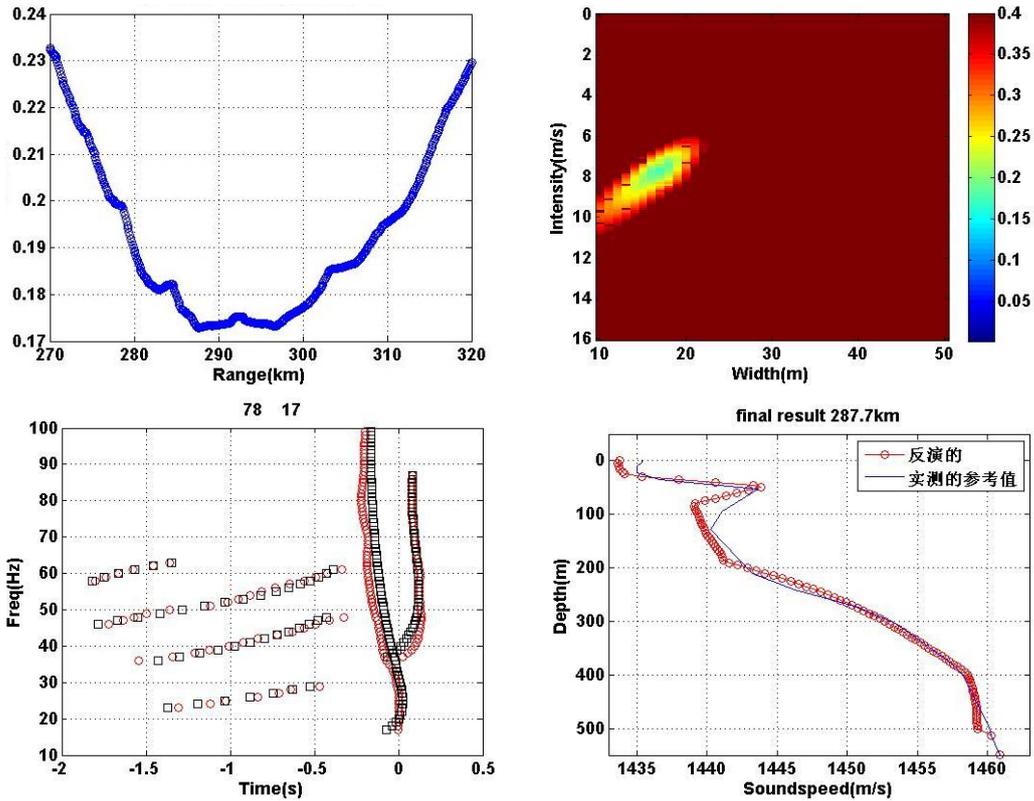

Figure 17 Cost function ambiguity map and sound velocity profile for 300km signal propagation inversion

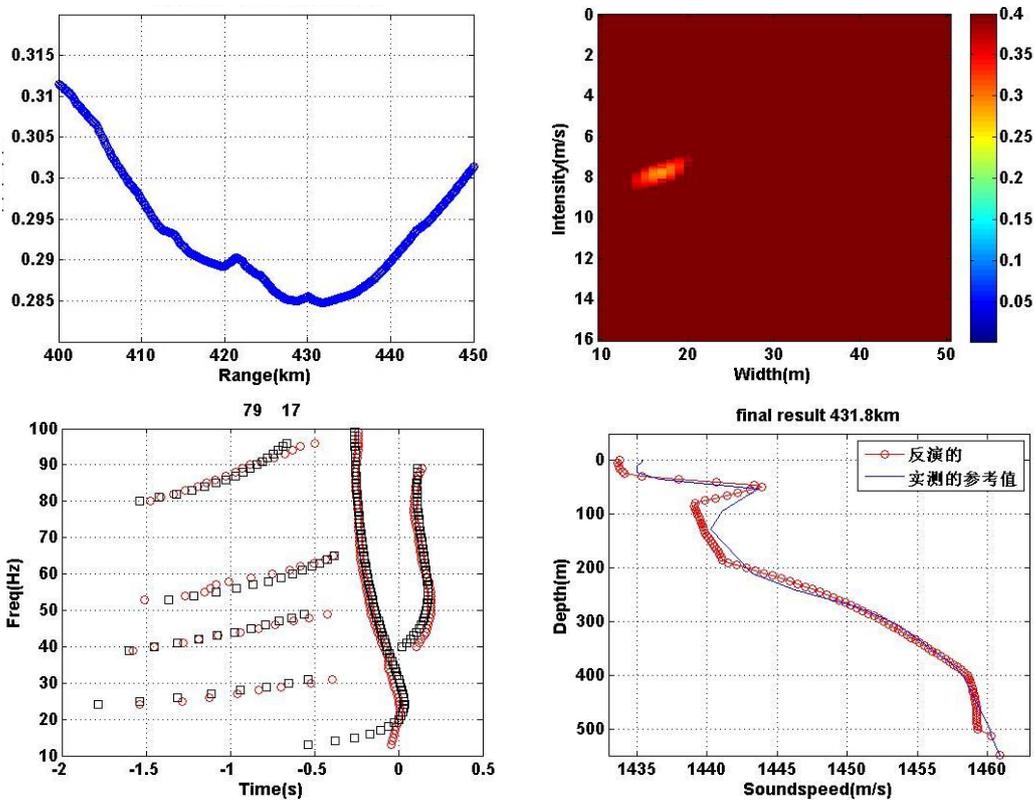

Figure 18 Cost function ambiguity map and sound velocity profile for 400km signal propagation inversion

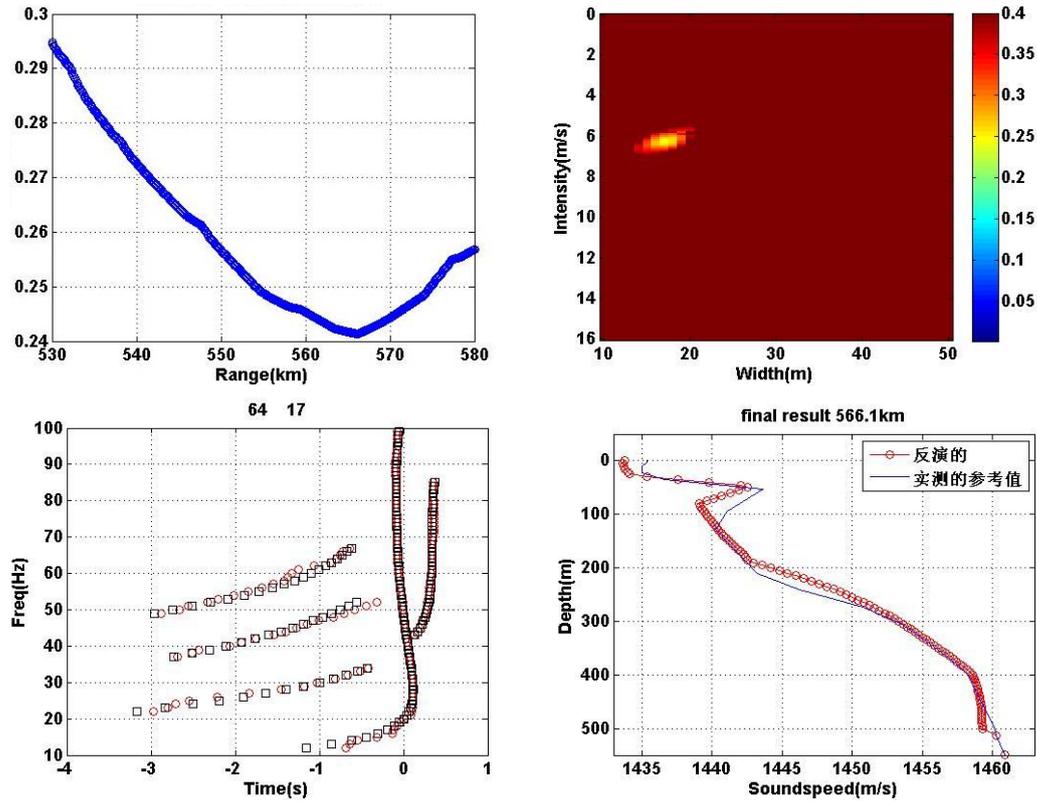

Figure 19 Cost function ambiguity map and sound velocity profile for 518km signal propagation inversion

Assuming the horizontal distance between the sound source and the receiver is known, the propagation distances are fixed at 200km, 300km, 400km, and 518km. In practical scenarios, the sound source location information can be obtained using AIS data, so the sound source distance information can be determined in advance during the inversion process, thereby further improving the accuracy of the inversion results. The sound speed profile inversion results at the four distances are presented in the following four figures. It can be observed that good sound speed profile inversion results can still be obtained under the premise of known sound source distance. The dispersion curves calculated based on the inversion results are also consistent with the actual dispersion curves. However, when compared with the inversion results for unknown sound source distances, it is found that the sound speed profiles inverted under known distances exhibit more pronounced spatial variations. Specifically, the dual-channel sound speed profile inverted at 518km is significantly weaker than that inverted at 200km. This is more in line with the actual situation, as the sound propagation path at 518km has higher latitude and a weaker dual-channel sound speed profile. In summary, although the sound speed profile can still be inverted for unknown sound source distances, known sound source distances can enhance the effectiveness of the dual-channel sound speed profile inversion results. This is mainly due to the coupling between sound source distance and dispersion structure, where unknown sound source distances increase the uncertainty of sound speed profile inversion. Of course, if dispersion curves of more modes can be obtained, the effectiveness of the inversion results should be further improved. In conclusion, fully utilizing the distance information of random sound sources can effectively enhance the effectiveness of dual-channel sound speed profile inversion results.

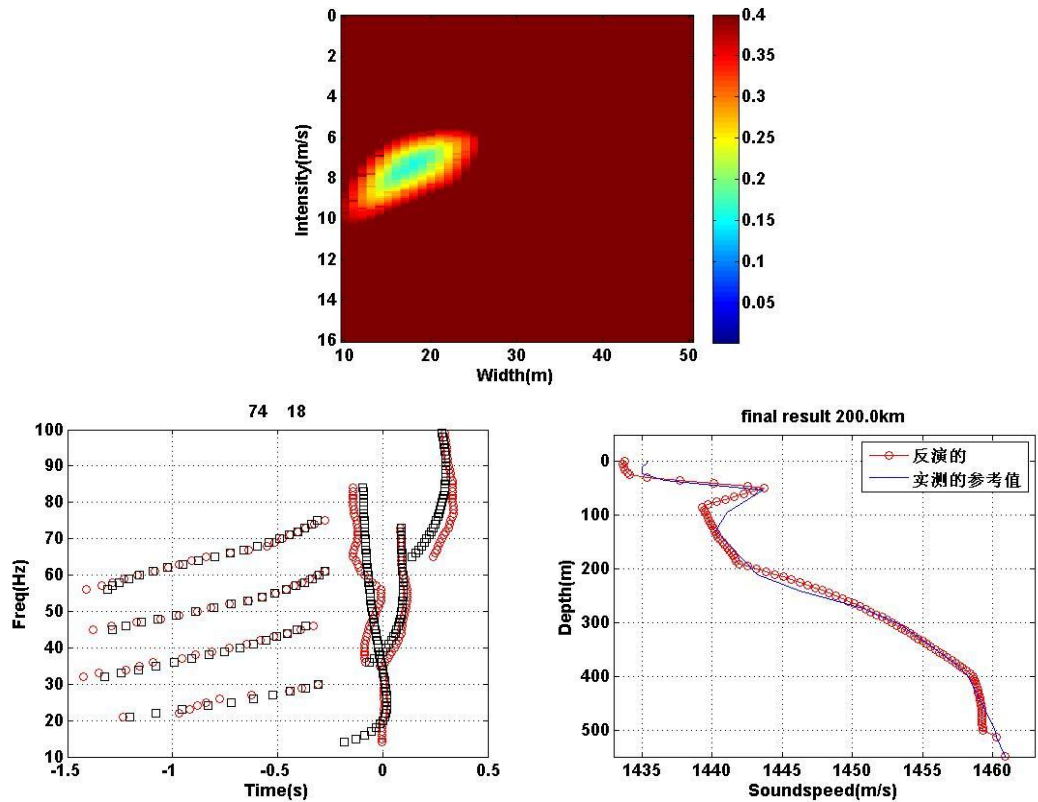

Figure 20 Cost function ambiguity diagram and sound speed profile for 200km propagation signal inversion (assuming the source distance is 200km)

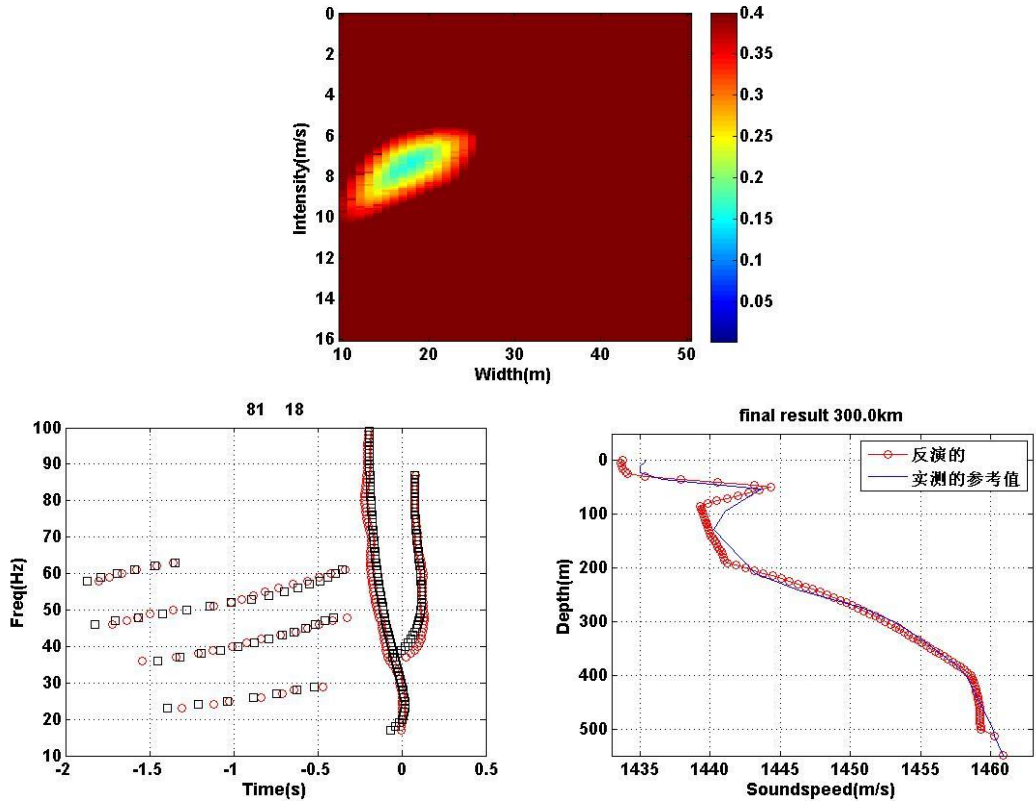

Figure 21 Cost function ambiguity diagram and sound speed profile for 300km signal propagation inversion (assuming the source distance is 300km)

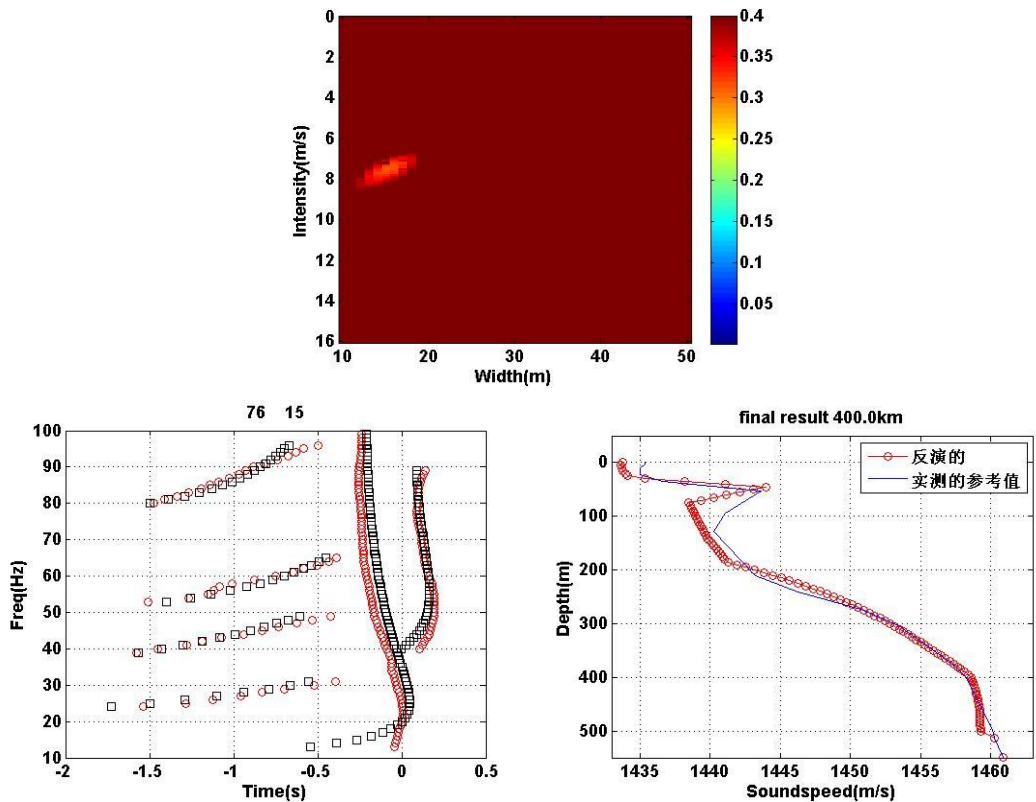

Figure 22 Cost function ambiguity diagram and sound speed profile for 400km signal propagation inversion (assuming the source distance is 400km)

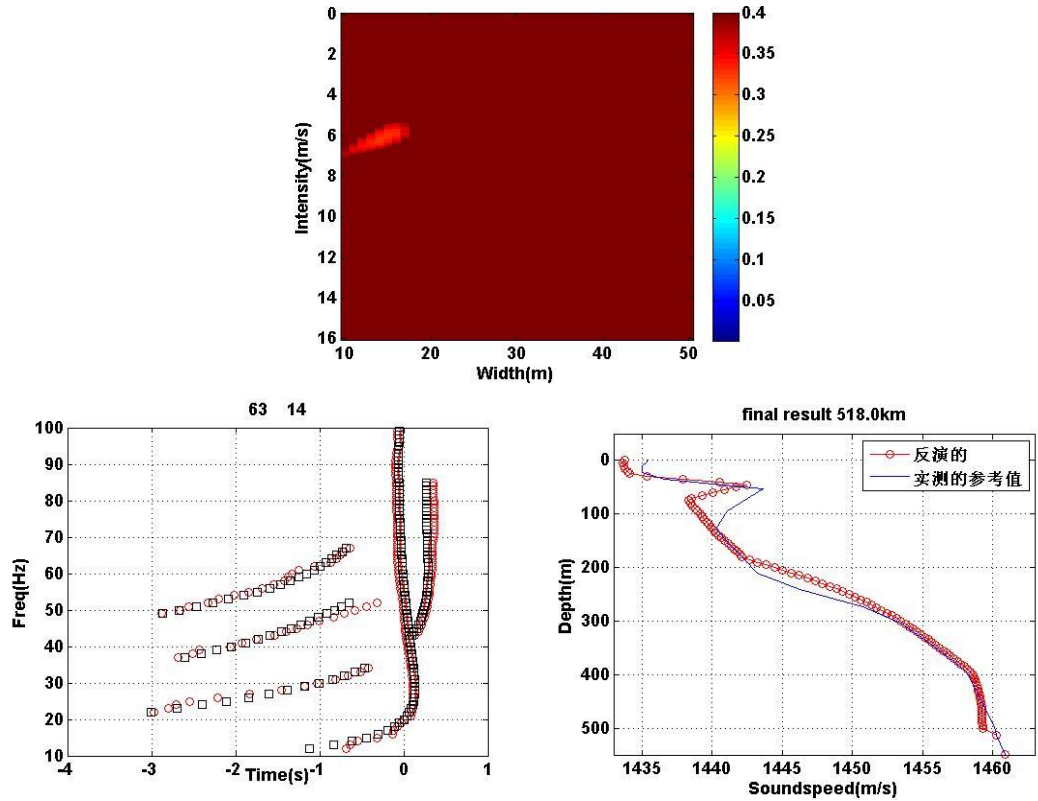

Figure 23 Cost function ambiguity map and sound speed profile for the inversion of a 518km propagated signal (assuming the source distance is 518km)

## B. Inversion result in range-dependent environment

For actual long-distance acoustic propagation routes, the sound speed profile often exhibits significant horizontal variations, and the inversion of horizontally varying sound speed profiles has always been a challenging problem in underwater acoustics research. To address this issue, this article utilizes the dispersion structure of acoustic signals at different distances to segmentally invert the two-channel sound speed profiles at different distances. The figure below presents the inverted sound speed profiles at four distances, as well as the two-dimensional sound speed profile field obtained using horizontal linear interpolation. This is the inversion result after dividing into four distance segments, including 0km to 200km, 200km to 300km, 300km to 400km, and 400km to 518km. From these four inversion results, it can be observed that both the intensity and width of the two-channel sound speed profile exhibit significant variation trends with distance. As can be seen from the figure, the inverted sound speed profile well reflects the characteristic that the two-channel sound speed profile gradually decreases with increasing distance, which is consistent with the actual horizontal variation of the two-channel sound speed profile. Therefore, the segmented two-channel sound speed profile inversion method proposed in this article is effective.

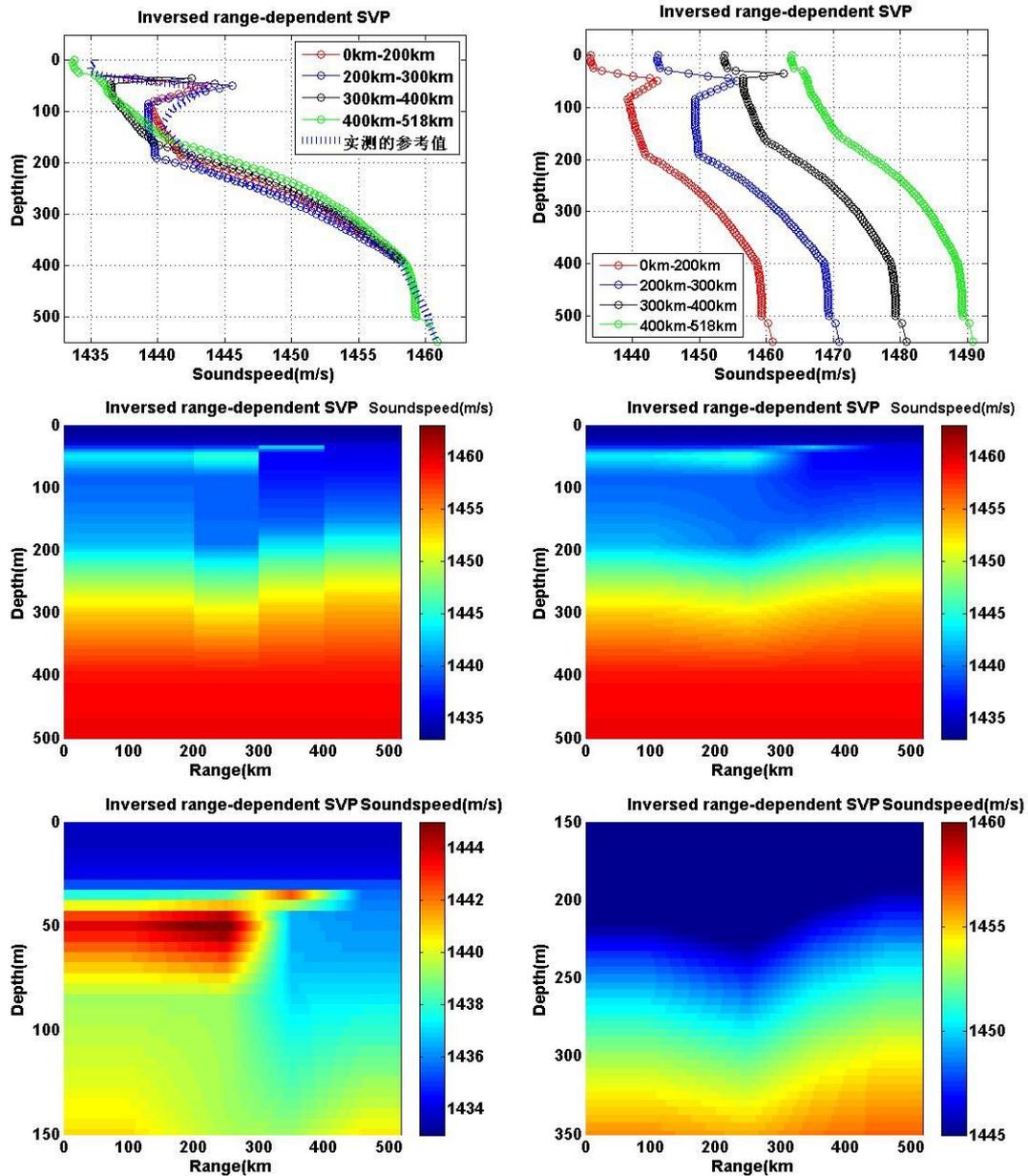

Figure 24 Comparison of sound speed profiles at different distance ranges, and a two-dimensional distribution diagram of how the sound speed profile varies with horizontal distance

Next, to further verify the effectiveness of the horizontal segmented inversion method, we further analyzed the inversion results of each segment, as shown in the figure below. From the inversion results of each segment, it can be observed that the dual-channel dual-parameter inversion ambiguity has good local optimal solutions. The inverted sound velocity profile obtained from the dispersion curve is also basically consistent with the measured dispersion curve, and the inverted dual-channel sound velocity profile is also relatively consistent with the measured dual-channel sound velocity profile. Furthermore, it shows a phenomenon of gradual decrease with increasing horizontal distance, which is consistent with the actual horizontal variation. Therefore, the method of using a segmented approach to invert the horizontally varying dual-channel sound velocity profile is effective.

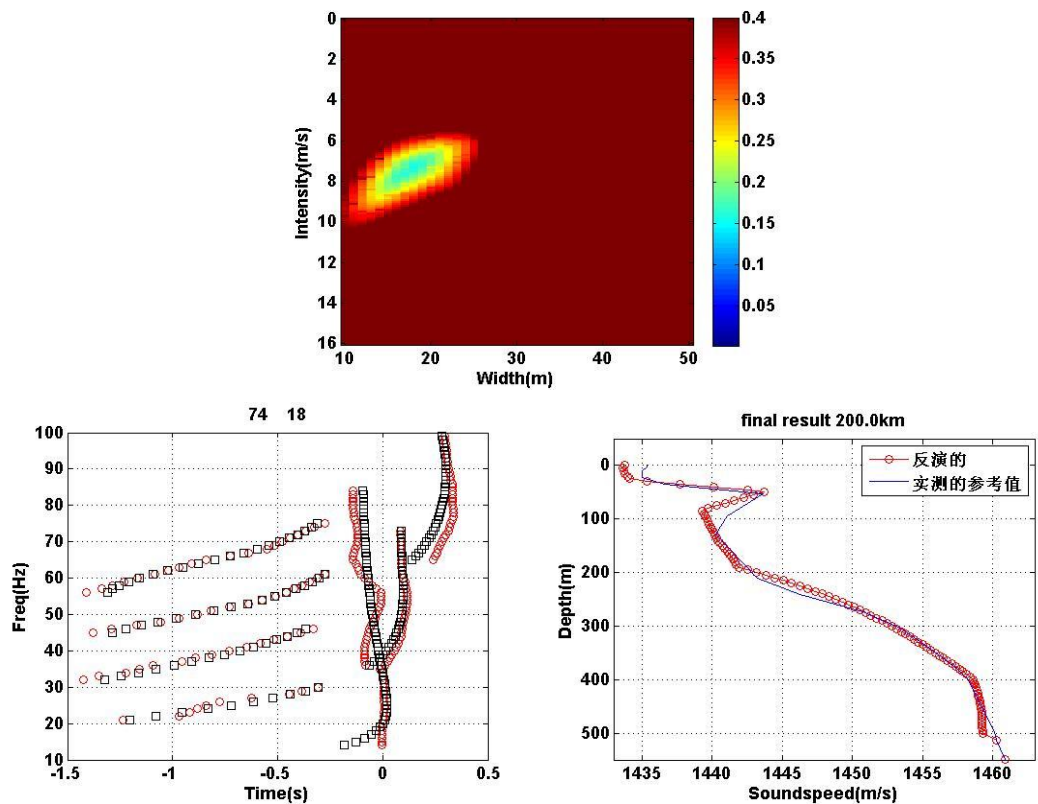

Figure 25 The inversion results from 000km to 200km, including the two-dimensional ambiguity map and sound speed profile

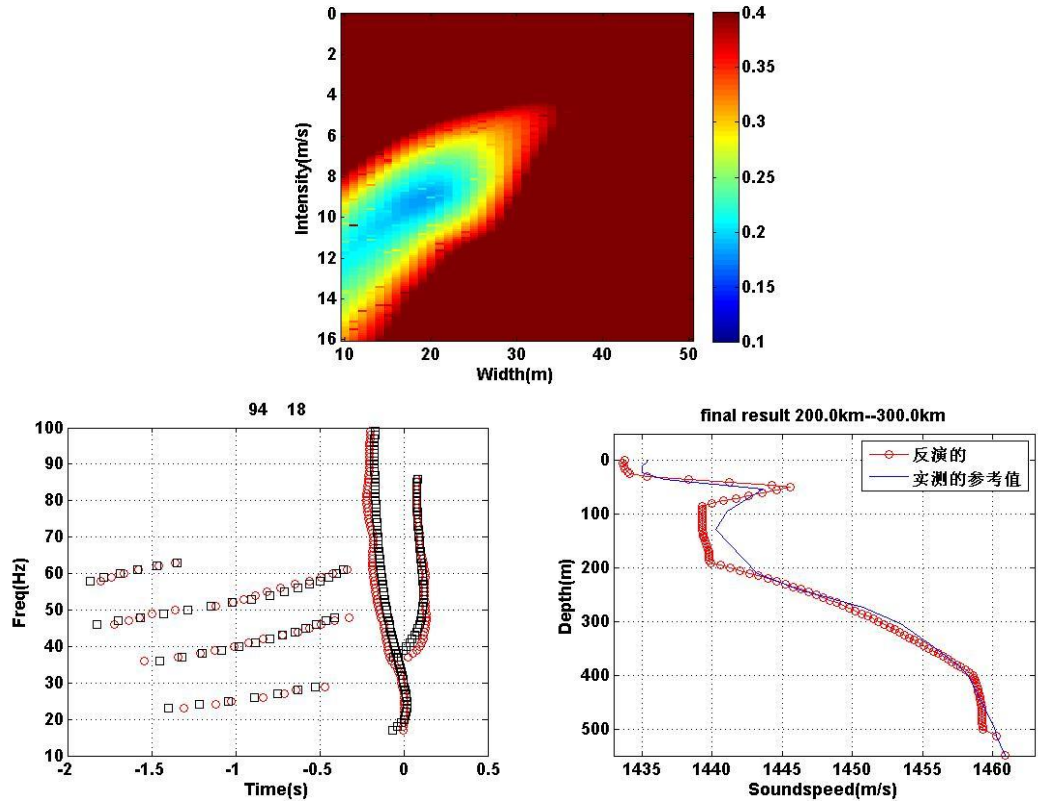

Figure 26 The inversion results from 200km to 300km, including the two-dimensional ambiguity map and sound speed profile

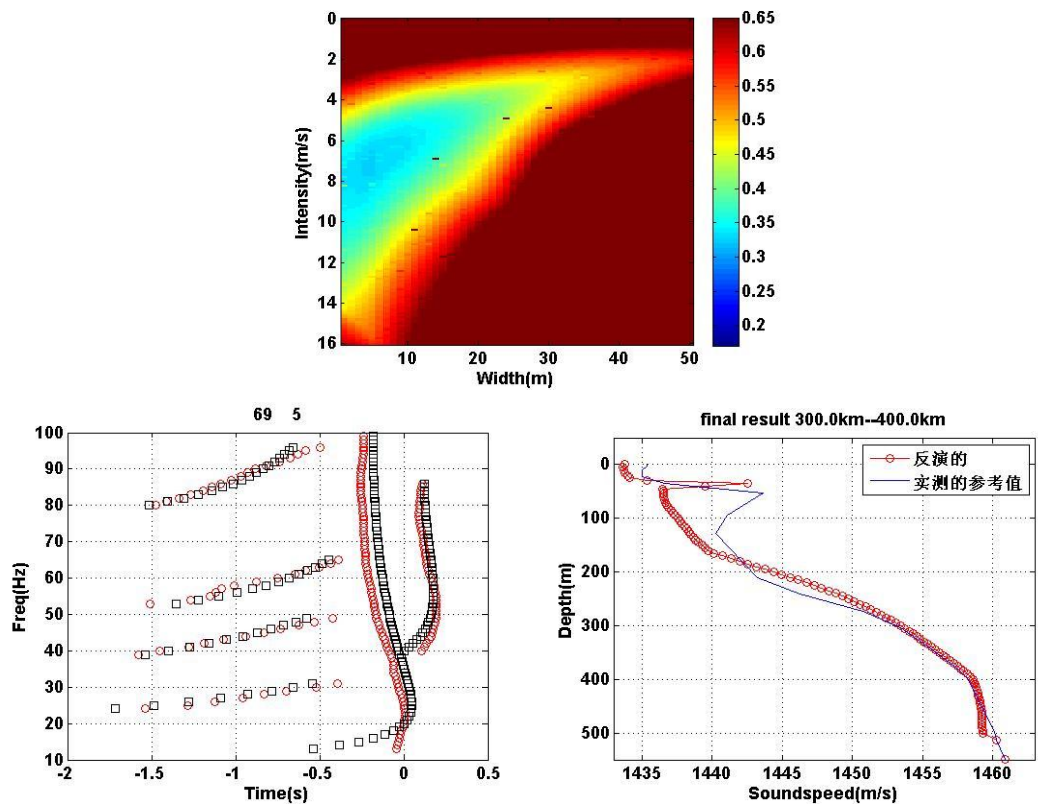

Figure 27 The inversion results from 300km to 400km, including the two-dimensional ambiguity map and sound speed profile

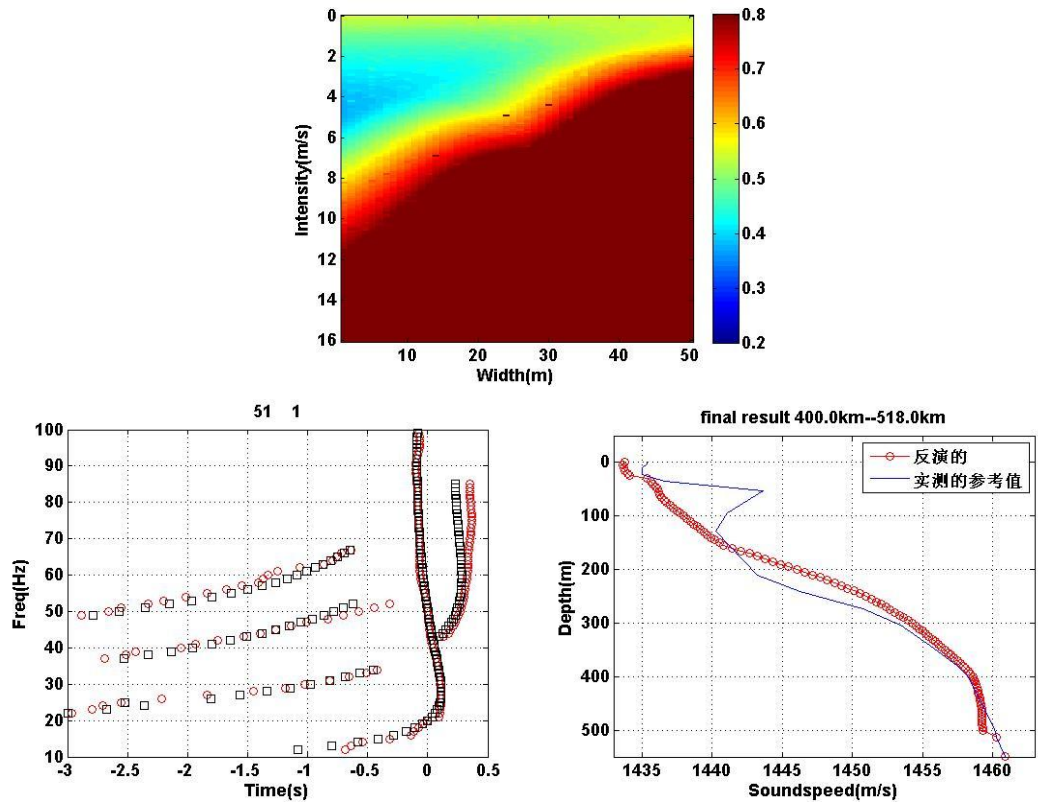

Figure 28 The inversion results from 400km to 518km, including the two-dimensional ambiguity map and sound velocity profile

## VII. CONCLUSION

This paper focused on the double-channel sound speed profile. For the recent decades, because of the intrusion of Pacific Ocean warm water, there are widespread double-channel sound speed profile in Chukchi Plateau and Canada Basin. Based on the recent research on the double-channel SSP, the double-channel SSP has a great potential in long-range detection and communication. However, recently, there are still quite few ways for wide-spread double-channel SSP measurement. This paper proposed a double-channel SSP inversion theory based on single hydrophone receiving random air-gun signal. Compared with traditional inversion theory, this inversion theory has the advantage of low-cost, easy-deployed and fast inversion. In this paper, the double-channel SSP was inversed in a range-independent environment, and also inversed in a range-dependent environment, and these two cases were verified by experiment datasets. Interestingly, this paper used random air-gun signals for SSP inversion, which means that the inversion need very-low cost, and this method was an example of low-cost ocean environment sound inversion. Besides, due to the shallow depth of air-gun source, the air-gun signal could extract the shallow water's environment information.

Another innovation of this paper is that the inversion method didn't need to know the propagation range under the lack of the exact propagation time, and this paper used the relative dispersion structure to inverse the double-channel SSP. And, for the double-channel SSP, the SSP could not be inversed by only used the low-frequency and non-intersecting portion of dispersion structure, because the high-frequency part of normal mode was concentrated on the shallow water, which means the high-frequency part carried more information about the double-channel SSP. This paper gave a good example for the inversion of double-channel SSP in Arctic.

Under the typical SSP in Arctic, the sound speed will increase with the depth in the surface water of deep sea in Arctic, and the surface water will from a refracted normal mode waveguide based on normal mode theory. And for the refracted normal mode, the normal mode dispersion line will increase with the time and frequency. However, under the dual-channel SSP, the normal mode will from an intersection of dispersion curves due to the nonlinear increasing sound velocity profile. After analysis, the intersection of refracted normal mode dispersion curves could be affected by the strength and width of the dual-channel SSP. Based on this principle, this paper proposed a dual-channel SSP inversion method based on the intersection phenomenon of refracted normal mode dispersion curves.

For the intersection phenomenon of normal mode dispersion curves under the dual-channel SSP, the intersected dispersion curves could not be easily extracted. While for the normal non-intersected dispersion curves, the dispersion curves could be extracted by refracted normal mode warping processing. However, for the dual-channel SSP, the dispersion curves could be extracted in two steps. Firstly, the intersected dispersion curves could be separated into two parts, lower frequency part and higher frequency part. Secondly, the lower part of dispersion curves could be extracted by refracted normal mode warping processing, because the lower part of dispersion curves didn't have the intersection part. Thirdly, the higher part of the dispersion curves could be extracted by extracting the energy peak time under different frequency. Finally, the complete dispersion curves could be achieved by combing the lower part and the higher part.

For the dual-channel SSP in Chukchi Plateau and Canada Basin, based on the widespread and large amount of SSP measurements, according to the analysis of the difference of the dual-channel SSP and the half waveguide in central Arctic, a dual-parameter representation method for the dual-channel SSP was proposed, and these two parameters represent the strength and width of the dual-channel SSP.

Based on the intersection phenomenon and the dual-parameter representation of the dual-channel SSP, a fast inversion method for the dual-channel SSP was proposed. Under the lack knowledge of sound source range, this method can inverse the range of the sound source together as an unknown variable. For the range-dependent ocean environment, after achieving multiple sound signals at different distances, this method can perform segmented inversion to obtain the dual-channel SSP that varies with distance.

In order to verify the dual-channel SSP inversion method, this paper used a random air-gun signal dataset, which was very common in Arctic. This air-gun dataset was produced by a seismic survey in Chukchi Plateau, and the position information could be achieved by analysis the AIS data. After AIS data analysis, the distance between the air-gun source and the receiving equipment varied from about 100 km to 518 km in Chukchi Plateau, which provided a very good sound propagation data for the dual-channel SSP inversion. Due to the randomness of the received air-gun signal, the SSP in the experiment area wasn't measured in the experiment time. However, a history SSP data in the Chukchi Plateau was used for inversion result comparing.

After extracting the intersection dispersion curves of refractive normal mode from the random air-gun signals, the dual-channel SSP inversion was conducted in two ways, including the range-independent cases and the range-dependent case. For the range-independent cases, this inversion method could achieve a reasonable result compared with the measured dual-channel SSP. And for the range-dependent case, this inversion method could not only provide a reasonable result, but also could show the changing trend of the dual-channel SSP along with the latitude, which was agreed with the measured SSPs. By the random air-gun sound signal datasets, this dual-channel SSP inversion method's ability was verified in the range-independent and the range-dependent cases.

Based on the intersection phenomenon of refracted normal mode under the dual-channel SSP, this paper proposed a single-hydrophone dual-channel SSP inversion method based on the random received wide-band impulse sound signal, and a random air-gun signal dataset was used to verify this inversion method. Actually, for the normal half waveguide SSP, the inversion method based on the normal mode dispersion curves could also achieve a good result, because the normal half waveguide SSP could be very easy to describe by two parameters. Due the commonness of air-gun signal in Arctic, this method based on random signal has a great potential for future development. And this ocean environment research could also used in the observation of changing water in the Arctic. However, for the shallow water area in the Arctic, this inversion could be useless because the normal mode which was bottom-reflected has a different dispersion curve.

## ACKNOWLEDGEMENTS

This research was supported in part by the National Key Research and Development Program of China under Grant, in part by the National Natural Science Foundation of China under Grant, in part by the Scientific Research Foundation of Third Institute of Oceanography, Ministry of Natural Resources under Grant, and in part by the Opening Foundation of State key laboratory of acoustics under Grant. Special thanks to all the staff involved in the deployment of acoustic equipment during the 11th Arctic voyage in 2020 and the recycling of acoustic equipment during the 12th Arctic voyage in 2021.